\documentclass[%
 reprint,
nofootinbib,
 amsmath,amssymb,
 aps,
prl,
]{revtex4-1}

\usepackage{xcolor}

\usepackage{graphicx}
\usepackage{dcolumn}
\usepackage{bm}

\usepackage{color,ulem}

\newcommand{\Ca}{\text{Ca}}

\begin{document}

\preprint{XXX}

\title{Viscous resuspension of droplets}

\author{Mehdi Maleki}
\author{Clément de Loubens }%
 \email{clement.de-loubens@univ-grenoble-alpes.fr}
 \author{Hugues Bodiguel}
 \affiliation{Univ. Grenoble Alpes, CNRS, Grenoble INP, LRP, 38000 Grenoble, France}

\date{\today}

\begin{abstract}
Using absorbance measurements through a Couette cell containing an emulsion of buoyant droplets, volume fraction profiles are measured at various shear rates. These viscous resuspension experiments allow a direct determination of the normal stress in the vorticity direction in connection with the suspension balance model that has been developed for suspensions of solid particles. The results unambigously show that the normal viscosity responsible for the shear-induced migration of the droplets is independant on the capillary number, implying that particle deformation does not play a great role. It is similar to that of rigid particles at volume fractions below 40\% but much smaller at higher ones.


\end{abstract}

\maketitle

 
Inspired by margination in blood vessels \citep{fahraeus1931viscosity,munn2008blood,pries1990blood}, there is a growing interest for flow-induced structuration phenomena in suspensions of soft microparticles in order to design cells sorting microfluidic systems for biological analysis \citep{goldsmith1984margination,shevkoplyas2005biomimetic,henry2016sorting}.  Understanding and modelling migration in suspensions of soft particles is considered a veritable challenge because of the non-linear coupling between hydrodynamic interactions and the dynamics of deformation of the particles \citep{kumar2012margination, freund2014numerical}. In fact, the deformability of the particle is a major ingredient to break the symmetry of inertialess flow. It generates normal forces and particle migration across the streamlines \citep{callens2008hydrodynamic, kaoui2008lateral}. The role of soft lubrication interactions between  particles has also been questioned as a collective mechanism of  migration \citep{gires2014pairwise,grandchamp2013lift}. 
Up to now, the problem has been addressed by full numerical simulations  which consider the interplay between the flow, the dynamics of deformation of the particles and their mechanical properties \citep{bagchi2010rheology,clausen2011rheology,zhao2012shear,fedosov2014white,freund2014numerical}. Even in a simple flow, the dynamic of deformation of soft particles is the complex result between the nature of the flow and the mechanical properties of the particle \citep{deschamps2009phase, de2016tank}. These complex dynamics would seem to play a role in particle migration \cite{malipeddi2021shear}. In contrast, this paper shows that all these sophisticated details do not need to be considered to account for migration in suspension of particles in a large range of volume fractions and shear rates.

The limiting case of rigid particles, where pair trajectories are perfectly reversible is worth discussion \cite{dacunha1996shear, chevremont2020lubricated}. Irreversibility has been discussed to be due to many body hydrodynamic interactions \cite{marchioro2001shear,sierou2004shear,pine2005chaos} but is now recognized to be more likely due to solid frictional contacts between rough particles \cite{dacunha1996shear, popova2007interaction,blanc2011experimental, pham2015particle,gallier2014rheology}. These  are thus  likely to play a major role in the shear-induced migration observed in heterogeneous flows of suspensions  \cite{lyon1998experimental,boyer2011dense}. Recent experimental results on viscous resuspension \cite{saint2019x,dambrosio2021} support this idea since migration was found to vary non-linearly with respect to the shear rate, which has been interpreted as a manifestation of non-Coulombian friction. Numerical investigations also highlighted the role of frictional contacts on the rheological properties of suspensions, but only at high volume fraction, since lubrication interactions are dominate for intermediate volume fraction  \cite{gallier2014rheology,mari2014shear,chevremont2019quantitative}. The exact role of contact contribution on particle migration thus remains to be clarified, and has important theoretical implications~\cite{lhuillier2009migration,nott2011suspension}. When reducing particle stiffness, it is expected that solid contacts between particles will eventually be precluded by a lubrication film. Therefore, studying shear-induced migration of deformable particles should also shed some lights on its physical origin in suspensions of rigid particles. A transition from a contact driven migration to a deformability driven one could thus be expected. It has been evidenced for pair trajectories \cite{chevremont2020lubricated}, but remains to be investigated for suspensions of soft particles.

\begin{figure*}
    \centering
    \includegraphics[width=\linewidth]{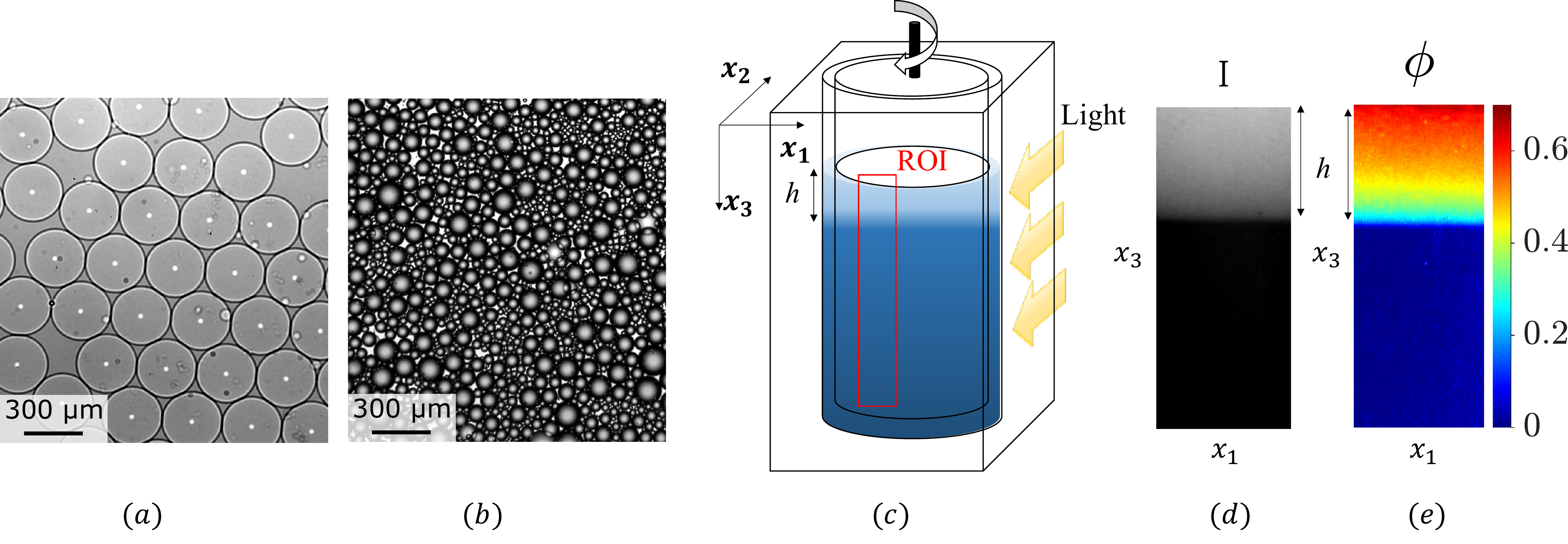}
    
    \caption{(a) Monodisperse and (b) polydisperse suspensions of oil-in-water droplets. (c) The suspension was sheared between the two concentric cylinders of a Taylor-Couette cell and visualized by light transmission in the region of interest (ROI) according to $\boldsymbol{x_2}$. $h$ is the height of the resuspended suspension. (d) Transmitted light intensity $I$ in the ROI. (e) Volume fraction of the droplets $\phi \propto \log_{10}(I_0/I)$ in the ROI.}
    
    \label{FigMethode}
\end{figure*}

In order to shed light on the collective mechanisms governing the migration of soft particles, we experimentally studied viscous resuspension of buoyant droplets, which are the simplest model of deformable particles, in a Taylor-Couette geometry, Fig. \ref{FigMethode}-c. In this configuration, migration occurs in the vorticity direction and is mainly the result of collective interactions between the droplets without interfering with other migration mechanisms such as wall effects and gradients of shear rate. An elegant way of interpreting migration in flow of suspension is to refer to a two-phase momentum balance model, the suspension balance model (SBM) \cite{nott1994pressure,morris1999curvilinear}, which had been developed for rigid particles. The interest of the SBM, compared to diffusive models \cite{leighton1986viscous,phillips1992constitutive,acrivos1993shear}, that it connects the particle migration to the suspension rheology. More specifically, the flux of particles is given by a momentum balance between the divergence of the particle stress tensor, buoyancy and the viscous drag. For rigid particles, constitutive relations between particle stress, volume fraction and shear rate have been determined \cite{lyon1998experimental,snook2016dynamics,boyer2011dense,leighton1986viscous, acrivos1993shear, zarraga2000characterization, saint2019x,dambrosio2021, boyer2011unifying,zarraga2000characterization}. For emulsions, earlier theoretical work \cite{Brenner,ramachandran2010constitutive} predicted  that migration and normal stress depends on both  volume fraction and capillary number $\Ca = \eta_0 \dot{\gamma} a/\sigma $  ($a$ being the droplet radius, $\eta_0$ the suspending liquid viscosity and $\sigma$ the surface tension), which is verified in the dilute case  \cite{king2001measurement}. However, there is a serious lack of experimental data about droplet migration in semi-dilute regime, despite a few  observations suggesting that shear-induced migration in emulsions is effective \cite{hollingsworth2006droplet, abbas2017pipe}. Here, by filling this gap, we show that the SBM fully accounts for the viscous resuspension in emulsions, and unexpectedly, that the normal particle stress is linear with respect to the shear rate and independent of  $\Ca$, similarly to the rigid case.


Emulsions of non-Brownian droplets were made by dispersing a medium chain triglyceride oil (Nestlé, Switzerland) in an aqueous solution of glycerol (84 \% w/w, CAS number 56-81-5, VWR) and were stabilized with 1\% w/w sorbitan trioleate 85 (CAS number 26266-58-0, Sigma-Aldrich). Monodisperse emulsions were produced with a microfluidic T-junction, radius $a =$ 143 $\pm$ 5 $\mu$m (Fig. \ref{FigMethode}-a), whereas a membrane emulsification device \citep{kosvintsev2005liquid, maleki2021membrane} (Micropore LDC-1, Micropore Technologies Ltd, UK) was used to produced polydisperse suspensions, $\overline{a} =$ 47 $\mu$m (Fig. \ref{FigMethode}-b, see Supp. Mat.). The viscosity and density of both phases were measured with a rotational rheometer (DHR3, TA instruments) and a densimeter (DMA 4500M, Anton Paar) and were $\eta_{d}=$ 50 mPa.s, $\eta_{c} =$ 85 mPa.s and $\rho_{d} =$ 943.42 kg/m$^3$, $\rho_{c} =$ 1218.52 kg/m$^3$ at 23$^{\circ}$C, respectively. The surface tension $\sigma$ between the two phases was measured by the pendant drop method and was 5 mN/m. 

Resuspension experiments were carried out in a home-made transparent Taylor-Couette cell, 50 mm high and made from PMMA. (Fig. \ref{FigMethode} -c) driven by a DHR3 rheometer (TA instruments). The inner and outer radii of the cell were $R_1=$ 20 and $R_2=$ 24 mm, respectively. The gap was large enough to accommodate at least 10 droplets, while minimizing the variations of shear rate $\dot{\gamma}$. The suspension of droplets was poured in the cell and left to cream for several hours to obtain a layer of creamed droplets of height $h_0$.
The emulsion was sheared at different increasing steps of $\dot{\gamma}$ from 6  to 260 $s^{-1}$. For each step, we waited until that the concentration profile reached a steady state. The maximal value of $\dot{\gamma}$ was chosen to keep the Reynolds number $Re =\rho \Omega R_{1} (R_{2} - R_{1})/\eta_0$ sufficiently low to avoid the emergence of Taylor vortices, i.e. $\textrm{Ta} = Re^22(R_2-R_1)/(R_1+R_2) < $ 900. It also made it possible to avoid the secondary currents which arise at higher shear rates from the combination of the centrifugal force and buoyancy \cite{saint2019x}. This range of shear rate corresponds to a variation of $\Ca$ between 10$^{-3}$ and 0.4.  Beyond this value, break-up of droplets was expected \citep{bentley1986experimental}. To ensure the absence of droplet break-up and/or coalescence, we tested the repeatability of the measurement at the smaller shear rate. Coalescence was furthermore avoided using surface treatment of the cell. Although these precautions were sufficient for the two monodispere emulsions studied, we were only able to limit coalescence for the polydisperse emulsions. However, we  quantified it and corrected the $h_0$ values (see Supp. Mat.).      

The resuspension process was visualized with a color camera in a region of interest (ROI), which was centered along the axis of rotation of the inner cylinder (Fig. \ref{FigMethode}-c), see Supp. Mat for details. The concentration profile $\phi(x_3)$ was inferred by light absorption technique. The optical indexes of both phases of the emulsion were precisely matched. A non-fluorescent colorant (E122, Breton) was added to the continuous phase to provide light absorbance contrast (Fig. \ref{FigMethode}-d). $\phi = K A$ was then inferred from the measurement of the absorbance $A = \log_{10}(I_0/I)$ for each pixel, where $I$ is the light intensity, $I_0$ the intensity of the background and $K$ the coefficient of attenuation (Fig. \ref{FigMethode}-d,e).

\begin{figure}
	\centering
	\includegraphics[width=1\columnwidth]{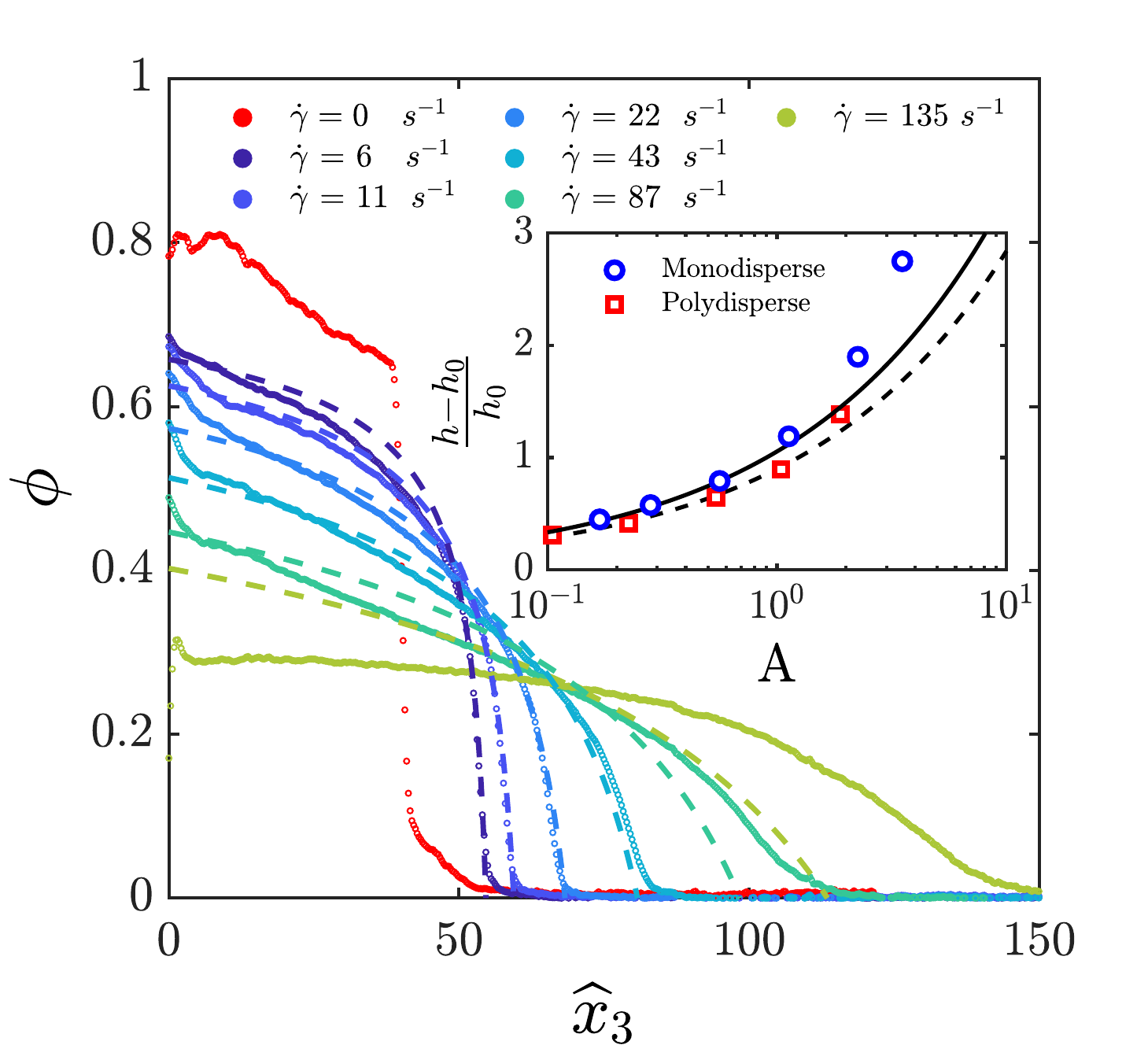}
	\caption{Steady state  concentration profiles $\phi(\widehat{x}_3)$ for each shear rate step  $\dot{\gamma}$ (colored dots) and Eq. \ref{Eq_phix3} with $\phi_m = 0.8$ and $\lambda_3 = $ 1 (colored dashed lines) for mono-dispersed suspension. In the insert, the relative height increment of the resuspended layer is plotted as a function of the Acrivos number, $A=a Sh/h_0$. The solid lines are the SBM predictions.}
	\label{Fif_ConcentrationProf}
\end{figure}


Fig. \ref{Fif_ConcentrationProf} shows the concentration profiles obtained on one of the emulsions used. At rest, the front of $\phi$ showed a sharp  transition between $\phi = $ 0 and 0.6. However, contrary to hard spheres, $\phi$  was not constant in the dense-packed zone and increased from 0.6 to 0.8. It is an indication of droplets deformation, because the droplets at the top of the layer underwent the hydrostatic pressure of the layers below \citep{henschke2002determination}. We defined a local Laplace number $La$, as the ratio of this pressure over the Laplace pressure, i.e. $La = a \int \phi \Delta \rho g dz/\sigma$ and showed (see Supp. Mat. ) that $\phi$ is, at rest, a unique function of $La$  up to 0.3.

\begin{figure*}
	\centering
	\includegraphics[width=1\linewidth]{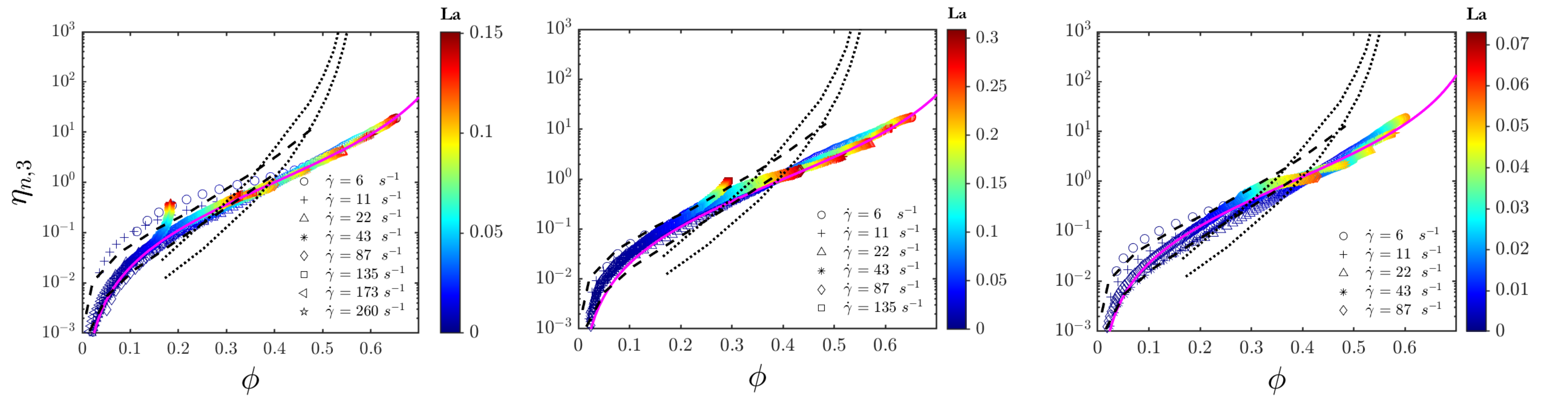}
	\caption{\small Relative normal viscosity $\eta_{n,3}$ as a function of the local droplet concentration $\phi$ for shear rates in the range of $\dot{\gamma} = 6$ to $260 \ s^{-1}$.  $\eta_{n,3}$ is inferred by integration of the concentration profile, Eq. \ref{IntPhi}. The solid lines are the best fit of Eq. \ref{Eq_phi_eta} (using $n=2$). \textbf{Left:} Monodisperse emulsion with $a=185\ \mu$m, $h_0 = 3.6$ mm, $\phi_m$ = 0.8 and $\lambda_3$ = 1 \textbf{Middle:} Monodisperse emulsion with $a=143\ \mu$m, $h_0 = 5.4$ mm, $\phi_m$ = 0.8 and $\lambda_3$ = 1 \textbf{Right:} Polydisperse emulsion with $\overline{a}=47\ \mu$m, $h_0 = 4$ to $6$ mm,  $\phi_m$ = 0.76 and $\lambda_3$ = 0.69. The color represents the Laplace number which denotes particle deformation. The dashed and dotted lines correspond to the experimental results on rigid particles of references \onlinecite{saint2019x} and \onlinecite{dambrosio2021}, respectively.}
		\label{FignetaP}
\end{figure*}

The increasing shear rate $\dot{\gamma}$, the height $h$ of the suspended layer increased as the mean value of $\phi$ decreased, due to volume conservation. Integration of $\phi$ showed that the volume of droplets was conserved throughout the experiment, which validated its measurement (see Supp. Mat.). In the SBM framework, the steady state volume fraction profiles result from the momentum balance in the particle phase, i.e. $\phi \Delta \rho g + \boldsymbol{\nabla}\cdot \boldsymbol{\Sigma}_p$. Under the hypothesis of linearity with respect to the shear rate, the particle normal stress could be written as $\Sigma_{p,ii} =  \eta_{0} \eta_{n,i}(\phi) \lvert \dot {\gamma} \rvert$ where $\eta_n$ is the non-dimensional normal viscosity \cite{nott1994pressure,morris1999curvilinear}. If the momentum balance is simplified, we easily obtain  
\begin{equation}
\label{Eq_phi_eta}
    \frac{\phi}{Sh}= -\frac{d\eta_{n,3}}{d\phi}\frac{d\phi}{d\widehat{x}_3}
\end{equation}
where $\widehat{x}_3=x_3/a$ and $Sh = \eta_0 \dot\gamma/\Delta \rho g a $ is the Shields number. Thus, integrating Eq. \ref{Eq_phi_eta} along the resuspended height provides a measurement of $\eta_{n,3}$ over a range of $\phi$ which depends on $Sh$ (or $\dot{\gamma}$)
\begin{equation}
\label{IntPhi}
    \eta_{n,3} = \frac{1}{Sh} \int_{\widehat{x}_3}^{h/a} \phi(u)du.
\end{equation}
The results of this integration are presented in Fig. \ref{FignetaP}. Unambiguously, most of the data fall on the same master-curve. Deviations observed for the smallest shear rates at low $\phi$ are not relevant since the corresponding volume fraction profiles tend towards zero very sharply, over a distance that is about the size of a droplet. For the highest shear rates, a small deviation could also be seen at the very top of the resuspended layer. It can have several origins (contribution of $\Ca$, inertia, radial migration ...).

For rigid particles, the normal viscosity is given by 
\begin{equation}
    \label{Eq_phi_eta2} 
    \eta_{n,i} = \lambda_i \phi^n (1-\phi/\phi_m)^{-n},
\end{equation}
where $\lambda_i$ are anisotropy coefficients, $\phi_m$ the maximal volume fraction, and where the exponent $n$ has been found to be 2  \cite{morris1999curvilinear,boyer2011dense,saint2019x} or 3  \cite{zarraga2000characterization,dambrosio2021}).  The evolution of $\eta_{n,3}(\phi)$ is very well fitted by Eq. \ref{Eq_phi_eta2} with $n =$ 2. The validity of these fits was also tested directly on $\phi(\widehat{x}_3)$ which are more sensitive to the values of the coefficients. For $n =$ 2, the concentration profile is given by \citep{saint2019x}
\begin{equation}
    \frac{\phi(\widehat{x}_3)}{\phi_m} = 1 - \left[ 1 + \frac{\phi_m}{\lambda_3Sh}(\widehat{h}-\widehat{x}_3)\right]^{-1/2} 
    \label{Eq_phix3}
\end{equation}
where the normalized height of the suspension in the steady state $\widehat{h}$ is
\begin{equation}
  \widehat{h} = \widehat{h}_0 + 2 \sqrt{\frac{\lambda_3 \widehat{h}_0}{\phi_m}Sh}. \label{eq_totalHeught}  
\end{equation}

The experimental and analytical profiles of $\phi(\widehat{x}_3)$  are in very good agreement (see Fig. \ref{Fif_ConcentrationProf}, and supp. mat.), except for the highest value of  $\dot{\gamma}$. The relative height increments, (insert in Fig. \ref{Fif_ConcentrationProf}) also show a very good agreement with Eq. \ref{eq_totalHeught}. 

These results shows that a simple constitutive equation of $\eta_{n,3}$ for a suspension of deformable droplets is sufficient to catch migration phenomena without further sophistication of the SBM. Strikingly, even when droplet deformation cannot be neglected at rest, i.e. for Laplace number above 0.05, it does not affect the normal viscosity under shear up to $La\sim 0.3$. The linear dependency with respect to the shear rate rules out some significant contribution of migration mechanisms due to drop deformability in the vorticity direction. This result is consistent with some simulation results showing a weak dependence of the normal forces when varying $Ca$ \cite{Aouane}.

The polydisperse emulsion exhibited a very similar behavior, with only small differences in the coefficients. This result is rather surprising as some size segregation of the droplet under shear could be expected. Our results indicate this kind of behaviour does not affect the macroscopic particle viscosity or the maximal volume fraction. 

The main difference with rigid particles concerns the value of $\phi_m$ which is comprised between 0.53-0.63, whereas it is 0.8 for the emulsions. As highlighted in Fig.~\ref{FignetaP}, the particle normal viscosity is therefore much higher for rigid ones than for droplets when $\phi > 0.4$. At these volume fractions, frictional contacts dominate the rheology of rigid ones \cite{chevremont2019quantitative} whereas droplets are likely to be frictionless.  
  
\begin{figure}
	\centering
	\includegraphics[width=1\columnwidth]{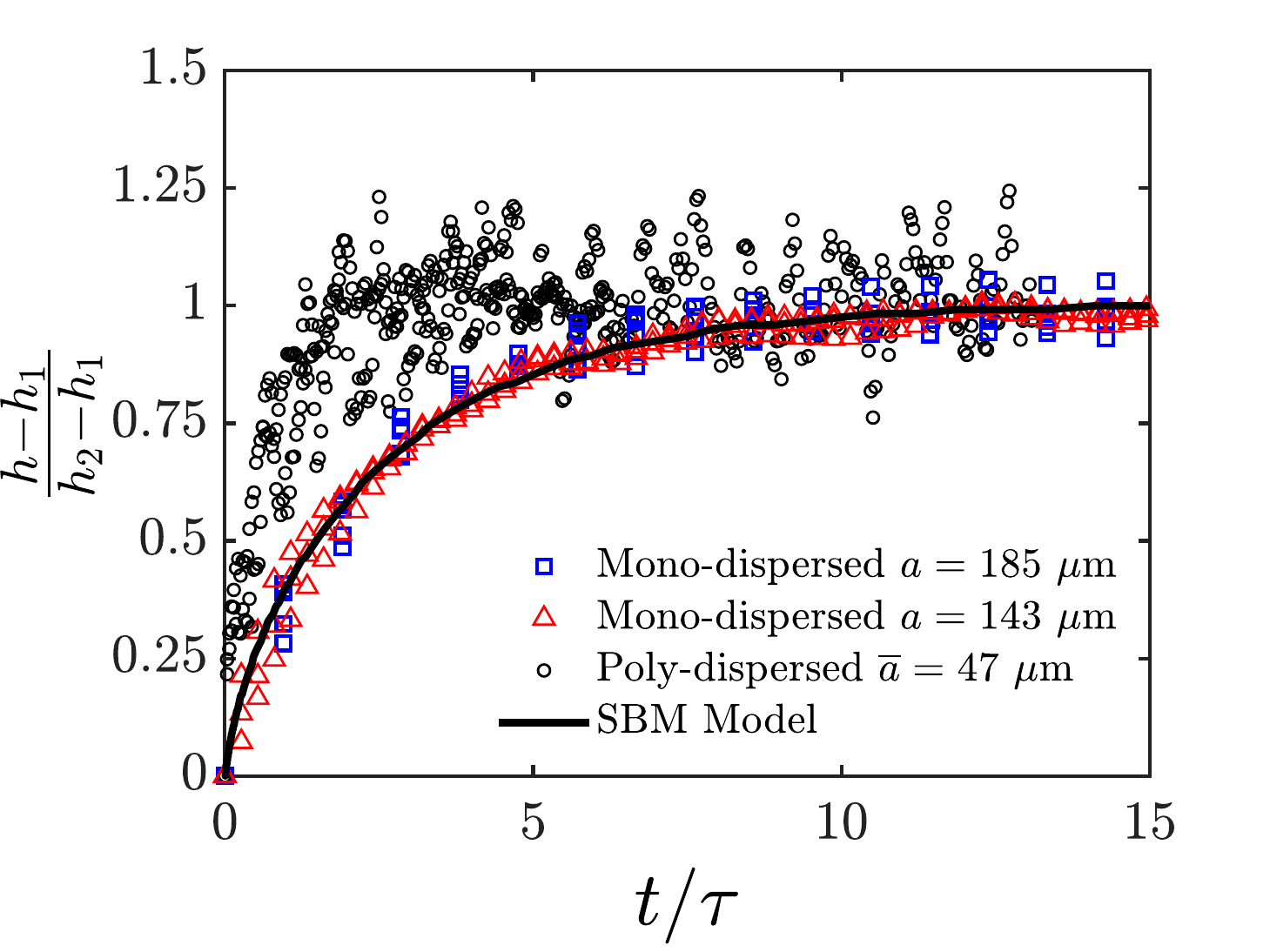}
	\caption{\small  Variation of the total height $h$ of resuspended layer, when the shear rate is suddenly changed from $\dot\gamma_1$ to $\dot\gamma_2$. The corresponding steady state heights are $h_1$ and $h_2$, respectively. The time is normalized by $\tau = \eta_0 h_0/\Delta \rho g a^2$. Several experiments are plotted together for each system ($\dot\gamma$ ranges from 10 and 100 s$^{-1}$), and collapse on a single curve. The solid lines are the calculated solutions of the SBM (see text).} 

	\label{Fig_dynamics}
\end{figure}

Let us now discuss the dynamics of resuspension. Momentum balance in the direction $x_3$ and mass conservation of the particulate phase read
 \begin{eqnarray}
     \frac{\partial\Sigma_{p,33}}{\partial x_3}-\frac{9}{2}\frac{\eta_0}{a^2}\frac{\phi}{f(\phi)}(u_{p,3}-u_{3})+\Delta \rho g \phi = 0 \\
\frac{\partial \phi}{\partial t} + \frac{\partial \phi u_{p,3}}{\partial x_3}=0 
     \label{Eq_dynamics} 
\end{eqnarray}
where $u_p$ and $u$ are the velocities of the particle phase and suspension phase, respectively. The second term in Eq. \ref{Eq_dynamics} corresponds to the viscous drag on the particle phase. $f$ is the hindered settling factor which is given by $(1-\phi)^5$ \citep{richardson1997sedimentation}. For suspensions of droplets, $f$ also depends on the viscosity ratio between the dispersed and continuous phases $\kappa$. Several expressions were proposed in Zinchenko \textit{et al.} \citep{zinchenko1984effect} and  Ramachandran \textit{et al.} \citep{ramachandran2010constitutive} in the limit of small $\phi$ and $Ca$. The simplest expression was $f_0=\left(2\kappa+2\right)/\left(9\kappa+6\right)$. We extend their results finite values of $\phi$ and $Ca$ by writing $ f= f_0 \left(1-\phi\right)^5 $. The system of equation \ref{Eq_dynamics} was solved for $u_3=0$ and using the normal viscosity determined in steady state. They are compared to the experimental results shown in Fig. \ref{Fig_dynamics}. An excellent agreement is found for monodisperse emulsions. In the range of parameters investigated, the kinetics is governed by a single characteristic time, $\tau = \eta_0 h_0/ \Delta \rho g a^2 $, and does not significantly depend on the shear rate.  For the polydisperse case, the kintetics is about two times faster, but this is not surprising as it is very sensitive to the particle size.


We conclude that the SBM quantitatively accounts for the viscous resuspension of droplets. Moreover, the particle normal stress is linear with respect to the shear rate, and proportional to $\phi^2/(1-\phi/\phi_m)^2$, similarly to rigid particles but with a higher maximal volume fraction. This result has important consequences. The first is the role of the droplet deformability. The normal stress, which is in principle a function of $\phi$ and $\Ca$,  does not depend on $\Ca$ up to 0.4. This implies that in the vorticity direction collective effects prevails over the coupling between particle shape and flow. 
The second consequence relates to the role of contact forces in shear-induced migration of particles. Leaving aside the small shear-thinning that has recently been reported for suspension of rigid particles \cite{saint2019x,dambrosio2021}, it is striking to observe that switching from rigid particles where frictional contacts dominate the rheology at high volume fractions, to droplets which could be considered as a frictionless particles, the normal stress is greatly reduced for $\phi>40\%$. This highlights the crucial role of frictional contacts in dense suspensions. Outlooks of this work deal with size segregation that should occur is the polydisperse case. 


\begin{acknowledgments}
LRP is part of the LabEx Tec21 (ANR-11-LABX-0030) and of the PolyNat Carnot Institute (ANR-11-CARN-007-01). The authors akcnowledge funding from Univ. Grenoble Alpes and from ANR (ANR-17-CE7-0040).  
\end{acknowledgments}

\bibliography{refs}

\begin{thebibliography}{56}%
\makeatletter
\providecommand \@ifxundefined [1]{%
 \@ifx{#1\undefined}
}%
\providecommand \@ifnum [1]{%
 \ifnum #1\expandafter \@firstoftwo
 \else \expandafter \@secondoftwo
 \fi
}%
\providecommand \@ifx [1]{%
 \ifx #1\expandafter \@firstoftwo
 \else \expandafter \@secondoftwo
 \fi
}%
\providecommand \natexlab [1]{#1}%
\providecommand \enquote  [1]{``#1''}%
\providecommand \bibnamefont  [1]{#1}%
\providecommand \bibfnamefont [1]{#1}%
\providecommand \citenamefont [1]{#1}%
\providecommand \href@noop [0]{\@secondoftwo}%
\providecommand \href [0]{\begingroup \@sanitize@url \@href}%
\providecommand \@href[1]{\@@startlink{#1}\@@href}%
\providecommand \@@href[1]{\endgroup#1\@@endlink}%
\providecommand \@sanitize@url [0]{\catcode `\\12\catcode `\$12\catcode
  `\&12\catcode `\#12\catcode `\^12\catcode `\_12\catcode `\%12\relax}%
\providecommand \@@startlink[1]{}%
\providecommand \@@endlink[0]{}%
\providecommand \url  [0]{\begingroup\@sanitize@url \@url }%
\providecommand \@url [1]{\endgroup\@href {#1}{\urlprefix }}%
\providecommand \urlprefix  [0]{URL }%
\providecommand \Eprint [0]{\href }%
\providecommand \doibase [0]{http://dx.doi.org/}%
\providecommand \selectlanguage [0]{\@gobble}%
\providecommand \bibinfo  [0]{\@secondoftwo}%
\providecommand \bibfield  [0]{\@secondoftwo}%
\providecommand \translation [1]{[#1]}%
\providecommand \BibitemOpen [0]{}%
\providecommand \bibitemStop [0]{}%
\providecommand \bibitemNoStop [0]{.\EOS\space}%
\providecommand \EOS [0]{\spacefactor3000\relax}%
\providecommand \BibitemShut  [1]{\csname bibitem#1\endcsname}%
\let\auto@bib@innerbib\@empty
\bibitem [{\citenamefont {Fahraeus}\ and\ \citenamefont
  {Lindqvist}(1931)}]{fahraeus1931viscosity}%
  \BibitemOpen
  \bibfield  {author} {\bibinfo {author} {\bibfnamefont {R.}~\bibnamefont
  {Fahraeus}}\ and\ \bibinfo {author} {\bibfnamefont {T.}~\bibnamefont
  {Lindqvist}},\ }\href@noop {} {\bibfield  {journal} {\bibinfo  {journal}
  {American Journal of Physiology-Legacy Content}\ }\textbf {\bibinfo {volume}
  {96}},\ \bibinfo {pages} {562} (\bibinfo {year} {1931})}\BibitemShut
  {NoStop}%
\bibitem [{\citenamefont {Munn}\ and\ \citenamefont
  {Dupin}(2008)}]{munn2008blood}%
  \BibitemOpen
  \bibfield  {author} {\bibinfo {author} {\bibfnamefont {L.~L.}\ \bibnamefont
  {Munn}}\ and\ \bibinfo {author} {\bibfnamefont {M.~M.}\ \bibnamefont
  {Dupin}},\ }\href@noop {} {\bibfield  {journal} {\bibinfo  {journal} {Annals
  of biomedical engineering}\ }\textbf {\bibinfo {volume} {36}},\ \bibinfo
  {pages} {534} (\bibinfo {year} {2008})}\BibitemShut {NoStop}%
\bibitem [{\citenamefont {Pries}\ \emph {et~al.}(1990)\citenamefont {Pries},
  \citenamefont {Secomb}, \citenamefont {Gaehtgens},\ and\ \citenamefont
  {Gross}}]{pries1990blood}%
  \BibitemOpen
  \bibfield  {author} {\bibinfo {author} {\bibfnamefont {A.~R.}\ \bibnamefont
  {Pries}}, \bibinfo {author} {\bibfnamefont {T.~W.}\ \bibnamefont {Secomb}},
  \bibinfo {author} {\bibfnamefont {P.}~\bibnamefont {Gaehtgens}}, \ and\
  \bibinfo {author} {\bibfnamefont {J.}~\bibnamefont {Gross}},\ }\href@noop {}
  {\bibfield  {journal} {\bibinfo  {journal} {Circulation research}\ }\textbf
  {\bibinfo {volume} {67}},\ \bibinfo {pages} {826} (\bibinfo {year}
  {1990})}\BibitemShut {NoStop}%
\bibitem [{\citenamefont {Goldsmith}\ and\ \citenamefont
  {Spain}(1984)}]{goldsmith1984margination}%
  \BibitemOpen
  \bibfield  {author} {\bibinfo {author} {\bibfnamefont {H.~L.}\ \bibnamefont
  {Goldsmith}}\ and\ \bibinfo {author} {\bibfnamefont {S.}~\bibnamefont
  {Spain}},\ }\href@noop {} {\bibfield  {journal} {\bibinfo  {journal}
  {Microvascular research}\ }\textbf {\bibinfo {volume} {27}},\ \bibinfo
  {pages} {204} (\bibinfo {year} {1984})}\BibitemShut {NoStop}%
\bibitem [{\citenamefont {Shevkoplyas}\ \emph {et~al.}(2005)\citenamefont
  {Shevkoplyas}, \citenamefont {Yoshida}, \citenamefont {Munn},\ and\
  \citenamefont {Bitensky}}]{shevkoplyas2005biomimetic}%
  \BibitemOpen
  \bibfield  {author} {\bibinfo {author} {\bibfnamefont {S.~S.}\ \bibnamefont
  {Shevkoplyas}}, \bibinfo {author} {\bibfnamefont {T.}~\bibnamefont
  {Yoshida}}, \bibinfo {author} {\bibfnamefont {L.~L.}\ \bibnamefont {Munn}}, \
  and\ \bibinfo {author} {\bibfnamefont {M.~W.}\ \bibnamefont {Bitensky}},\
  }\href@noop {} {\bibfield  {journal} {\bibinfo  {journal} {Analytical
  chemistry}\ }\textbf {\bibinfo {volume} {77}},\ \bibinfo {pages} {933}
  (\bibinfo {year} {2005})}\BibitemShut {NoStop}%
\bibitem [{\citenamefont {Henry}\ \emph {et~al.}(2016)\citenamefont {Henry},
  \citenamefont {Holm}, \citenamefont {Zhang}, \citenamefont {Beech},
  \citenamefont {Tegenfeldt}, \citenamefont {Fedosov},\ and\ \citenamefont
  {Gompper}}]{henry2016sorting}%
  \BibitemOpen
  \bibfield  {author} {\bibinfo {author} {\bibfnamefont {E.}~\bibnamefont
  {Henry}}, \bibinfo {author} {\bibfnamefont {S.~H.}\ \bibnamefont {Holm}},
  \bibinfo {author} {\bibfnamefont {Z.}~\bibnamefont {Zhang}}, \bibinfo
  {author} {\bibfnamefont {J.~P.}\ \bibnamefont {Beech}}, \bibinfo {author}
  {\bibfnamefont {J.~O.}\ \bibnamefont {Tegenfeldt}}, \bibinfo {author}
  {\bibfnamefont {D.~A.}\ \bibnamefont {Fedosov}}, \ and\ \bibinfo {author}
  {\bibfnamefont {G.}~\bibnamefont {Gompper}},\ }\href@noop {} {\bibfield
  {journal} {\bibinfo  {journal} {Scientific reports}\ }\textbf {\bibinfo
  {volume} {6}},\ \bibinfo {pages} {34375} (\bibinfo {year}
  {2016})}\BibitemShut {NoStop}%
\bibitem [{\citenamefont {Kumar}\ and\ \citenamefont
  {Graham}(2012)}]{kumar2012margination}%
  \BibitemOpen
  \bibfield  {author} {\bibinfo {author} {\bibfnamefont {A.}~\bibnamefont
  {Kumar}}\ and\ \bibinfo {author} {\bibfnamefont {M.~D.}\ \bibnamefont
  {Graham}},\ }\href@noop {} {\bibfield  {journal} {\bibinfo  {journal} {Soft
  Matter}\ }\textbf {\bibinfo {volume} {8}},\ \bibinfo {pages} {10536}
  (\bibinfo {year} {2012})}\BibitemShut {NoStop}%
\bibitem [{\citenamefont {Freund}(2014)}]{freund2014numerical}%
  \BibitemOpen
  \bibfield  {author} {\bibinfo {author} {\bibfnamefont {J.~B.}\ \bibnamefont
  {Freund}},\ }\href@noop {} {\bibfield  {journal} {\bibinfo  {journal} {Annual
  review of fluid mechanics}\ }\textbf {\bibinfo {volume} {46}},\ \bibinfo
  {pages} {67} (\bibinfo {year} {2014})}\BibitemShut {NoStop}%
\bibitem [{\citenamefont {Callens}\ \emph {et~al.}(2008)\citenamefont
  {Callens}, \citenamefont {Minetti}, \citenamefont {Coupier}, \citenamefont
  {Mader}, \citenamefont {Dubois}, \citenamefont {Misbah},\ and\ \citenamefont
  {Podgorski}}]{callens2008hydrodynamic}%
  \BibitemOpen
  \bibfield  {author} {\bibinfo {author} {\bibfnamefont {N.}~\bibnamefont
  {Callens}}, \bibinfo {author} {\bibfnamefont {C.}~\bibnamefont {Minetti}},
  \bibinfo {author} {\bibfnamefont {G.}~\bibnamefont {Coupier}}, \bibinfo
  {author} {\bibfnamefont {M.-A.}\ \bibnamefont {Mader}}, \bibinfo {author}
  {\bibfnamefont {F.}~\bibnamefont {Dubois}}, \bibinfo {author} {\bibfnamefont
  {C.}~\bibnamefont {Misbah}}, \ and\ \bibinfo {author} {\bibfnamefont
  {T.}~\bibnamefont {Podgorski}},\ }\href@noop {} {\bibfield  {journal}
  {\bibinfo  {journal} {EPL (Europhysics Letters)}\ }\textbf {\bibinfo {volume}
  {83}},\ \bibinfo {pages} {24002} (\bibinfo {year} {2008})}\BibitemShut
  {NoStop}%
\bibitem [{\citenamefont {Kaoui}\ \emph {et~al.}(2008)\citenamefont {Kaoui},
  \citenamefont {Ristow}, \citenamefont {Cantat}, \citenamefont {Misbah},\ and\
  \citenamefont {Zimmermann}}]{kaoui2008lateral}%
  \BibitemOpen
  \bibfield  {author} {\bibinfo {author} {\bibfnamefont {B.}~\bibnamefont
  {Kaoui}}, \bibinfo {author} {\bibfnamefont {G.}~\bibnamefont {Ristow}},
  \bibinfo {author} {\bibfnamefont {I.}~\bibnamefont {Cantat}}, \bibinfo
  {author} {\bibfnamefont {C.}~\bibnamefont {Misbah}}, \ and\ \bibinfo {author}
  {\bibfnamefont {W.}~\bibnamefont {Zimmermann}},\ }\href@noop {} {\bibfield
  {journal} {\bibinfo  {journal} {Physical Review E}\ }\textbf {\bibinfo
  {volume} {77}},\ \bibinfo {pages} {021903} (\bibinfo {year}
  {2008})}\BibitemShut {NoStop}%
\bibitem [{\citenamefont {Gires}\ \emph {et~al.}(2014)\citenamefont {Gires},
  \citenamefont {Srivastav}, \citenamefont {Misbah}, \citenamefont
  {Podgorski},\ and\ \citenamefont {Coupier}}]{gires2014pairwise}%
  \BibitemOpen
  \bibfield  {author} {\bibinfo {author} {\bibfnamefont {P.-Y.}\ \bibnamefont
  {Gires}}, \bibinfo {author} {\bibfnamefont {A.}~\bibnamefont {Srivastav}},
  \bibinfo {author} {\bibfnamefont {C.}~\bibnamefont {Misbah}}, \bibinfo
  {author} {\bibfnamefont {T.}~\bibnamefont {Podgorski}}, \ and\ \bibinfo
  {author} {\bibfnamefont {G.}~\bibnamefont {Coupier}},\ }\href@noop {}
  {\bibfield  {journal} {\bibinfo  {journal} {Physics of Fluids}\ }\textbf
  {\bibinfo {volume} {26}},\ \bibinfo {pages} {013304} (\bibinfo {year}
  {2014})}\BibitemShut {NoStop}%
\bibitem [{\citenamefont {Grandchamp}\ \emph {et~al.}(2013)\citenamefont
  {Grandchamp}, \citenamefont {Coupier}, \citenamefont {Srivastav},
  \citenamefont {Minetti},\ and\ \citenamefont
  {Podgorski}}]{grandchamp2013lift}%
  \BibitemOpen
  \bibfield  {author} {\bibinfo {author} {\bibfnamefont {X.}~\bibnamefont
  {Grandchamp}}, \bibinfo {author} {\bibfnamefont {G.}~\bibnamefont {Coupier}},
  \bibinfo {author} {\bibfnamefont {A.}~\bibnamefont {Srivastav}}, \bibinfo
  {author} {\bibfnamefont {C.}~\bibnamefont {Minetti}}, \ and\ \bibinfo
  {author} {\bibfnamefont {T.}~\bibnamefont {Podgorski}},\ }\href@noop {}
  {\bibfield  {journal} {\bibinfo  {journal} {Physical review letters}\
  }\textbf {\bibinfo {volume} {110}},\ \bibinfo {pages} {108101} (\bibinfo
  {year} {2013})}\BibitemShut {NoStop}%
\bibitem [{\citenamefont {Bagchi}\ and\ \citenamefont
  {Kalluri}(2010)}]{bagchi2010rheology}%
  \BibitemOpen
  \bibfield  {author} {\bibinfo {author} {\bibfnamefont {P.}~\bibnamefont
  {Bagchi}}\ and\ \bibinfo {author} {\bibfnamefont {R.~M.}\ \bibnamefont
  {Kalluri}},\ }\href@noop {} {\bibfield  {journal} {\bibinfo  {journal}
  {Physical Review E}\ }\textbf {\bibinfo {volume} {81}},\ \bibinfo {pages}
  {056320} (\bibinfo {year} {2010})}\BibitemShut {NoStop}%
\bibitem [{\citenamefont {Clausen}\ \emph {et~al.}(2011)\citenamefont
  {Clausen}, \citenamefont {Reasor},\ and\ \citenamefont
  {Aidun}}]{clausen2011rheology}%
  \BibitemOpen
  \bibfield  {author} {\bibinfo {author} {\bibfnamefont {J.~R.}\ \bibnamefont
  {Clausen}}, \bibinfo {author} {\bibfnamefont {D.~A.}\ \bibnamefont {Reasor}},
  \ and\ \bibinfo {author} {\bibfnamefont {C.~K.}\ \bibnamefont {Aidun}},\
  }\href@noop {} {\bibfield  {journal} {\bibinfo  {journal} {Journal of Fluid
  Mechanics}\ }\textbf {\bibinfo {volume} {685}},\ \bibinfo {pages} {202}
  (\bibinfo {year} {2011})}\BibitemShut {NoStop}%
\bibitem [{\citenamefont {Zhao}\ \emph {et~al.}(2012)\citenamefont {Zhao},
  \citenamefont {Shaqfeh},\ and\ \citenamefont {Narsimhan}}]{zhao2012shear}%
  \BibitemOpen
  \bibfield  {author} {\bibinfo {author} {\bibfnamefont {H.}~\bibnamefont
  {Zhao}}, \bibinfo {author} {\bibfnamefont {E.~S.}\ \bibnamefont {Shaqfeh}}, \
  and\ \bibinfo {author} {\bibfnamefont {V.}~\bibnamefont {Narsimhan}},\
  }\href@noop {} {\bibfield  {journal} {\bibinfo  {journal} {Physics of
  Fluids}\ }\textbf {\bibinfo {volume} {24}},\ \bibinfo {pages} {011902}
  (\bibinfo {year} {2012})}\BibitemShut {NoStop}%
\bibitem [{\citenamefont {Fedosov}\ and\ \citenamefont
  {Gompper}(2014)}]{fedosov2014white}%
  \BibitemOpen
  \bibfield  {author} {\bibinfo {author} {\bibfnamefont {D.~A.}\ \bibnamefont
  {Fedosov}}\ and\ \bibinfo {author} {\bibfnamefont {G.}~\bibnamefont
  {Gompper}},\ }\href@noop {} {\bibfield  {journal} {\bibinfo  {journal} {Soft
  matter}\ }\textbf {\bibinfo {volume} {10}},\ \bibinfo {pages} {2961}
  (\bibinfo {year} {2014})}\BibitemShut {NoStop}%
\bibitem [{\citenamefont {Deschamps}\ \emph {et~al.}(2009)\citenamefont
  {Deschamps}, \citenamefont {Kantsler},\ and\ \citenamefont
  {Steinberg}}]{deschamps2009phase}%
  \BibitemOpen
  \bibfield  {author} {\bibinfo {author} {\bibfnamefont {J.}~\bibnamefont
  {Deschamps}}, \bibinfo {author} {\bibfnamefont {V.}~\bibnamefont {Kantsler}},
  \ and\ \bibinfo {author} {\bibfnamefont {V.}~\bibnamefont {Steinberg}},\
  }\href@noop {} {\bibfield  {journal} {\bibinfo  {journal} {Physical review
  letters}\ }\textbf {\bibinfo {volume} {102}},\ \bibinfo {pages} {118105}
  (\bibinfo {year} {2009})}\BibitemShut {NoStop}%
\bibitem [{\citenamefont {de~Loubens}\ \emph {et~al.}(2016)\citenamefont
  {de~Loubens}, \citenamefont {Deschamps}, \citenamefont {Edwards-Levy},\ and\
  \citenamefont {Leonetti}}]{de2016tank}%
  \BibitemOpen
  \bibfield  {author} {\bibinfo {author} {\bibfnamefont {C.}~\bibnamefont
  {de~Loubens}}, \bibinfo {author} {\bibfnamefont {J.}~\bibnamefont
  {Deschamps}}, \bibinfo {author} {\bibfnamefont {F.}~\bibnamefont
  {Edwards-Levy}}, \ and\ \bibinfo {author} {\bibfnamefont {M.}~\bibnamefont
  {Leonetti}},\ }\href@noop {} {\bibfield  {journal} {\bibinfo  {journal}
  {Journal of Fluid Mechanics}\ }\textbf {\bibinfo {volume} {789}},\ \bibinfo
  {pages} {750} (\bibinfo {year} {2016})}\BibitemShut {NoStop}%
\bibitem [{\citenamefont {Malipeddi}\ and\ \citenamefont
  {Sarkar}(2021)}]{malipeddi2021shear}%
  \BibitemOpen
  \bibfield  {author} {\bibinfo {author} {\bibfnamefont {A.~R.}\ \bibnamefont
  {Malipeddi}}\ and\ \bibinfo {author} {\bibfnamefont {K.}~\bibnamefont
  {Sarkar}},\ }\href@noop {} {\bibfield  {journal} {\bibinfo  {journal} {Soft
  Matter}\ } (\bibinfo {year} {2021})}\BibitemShut {NoStop}%
\bibitem [{\citenamefont {Da~Cunha}\ and\ \citenamefont
  {Hinch}(1996)}]{dacunha1996shear}%
  \BibitemOpen
  \bibfield  {author} {\bibinfo {author} {\bibfnamefont {F.}~\bibnamefont
  {Da~Cunha}}\ and\ \bibinfo {author} {\bibfnamefont {E.}~\bibnamefont
  {Hinch}},\ }\href@noop {} {\bibfield  {journal} {\bibinfo  {journal} {Journal
  of fluid mechanics}\ }\textbf {\bibinfo {volume} {309}},\ \bibinfo {pages}
  {211} (\bibinfo {year} {1996})}\BibitemShut {NoStop}%
\bibitem [{\citenamefont {Ch{\`e}vremont}\ \emph {et~al.}(2020)\citenamefont
  {Ch{\`e}vremont}, \citenamefont {Bodiguel},\ and\ \citenamefont
  {Chareyre}}]{chevremont2020lubricated}%
  \BibitemOpen
  \bibfield  {author} {\bibinfo {author} {\bibfnamefont {W.}~\bibnamefont
  {Ch{\`e}vremont}}, \bibinfo {author} {\bibfnamefont {H.}~\bibnamefont
  {Bodiguel}}, \ and\ \bibinfo {author} {\bibfnamefont {B.}~\bibnamefont
  {Chareyre}},\ }\href@noop {} {\bibfield  {journal} {\bibinfo  {journal}
  {Powder Technology}\ }\textbf {\bibinfo {volume} {372}},\ \bibinfo {pages}
  {600} (\bibinfo {year} {2020})}\BibitemShut {NoStop}%
\bibitem [{\citenamefont {Marchioro}\ and\ \citenamefont
  {Acrivos}(2001)}]{marchioro2001shear}%
  \BibitemOpen
  \bibfield  {author} {\bibinfo {author} {\bibfnamefont {M.}~\bibnamefont
  {Marchioro}}\ and\ \bibinfo {author} {\bibfnamefont {A.}~\bibnamefont
  {Acrivos}},\ }\href@noop {} {\bibfield  {journal} {\bibinfo  {journal}
  {Journal of fluid mechanics}\ }\textbf {\bibinfo {volume} {443}},\ \bibinfo
  {pages} {101} (\bibinfo {year} {2001})}\BibitemShut {NoStop}%
\bibitem [{\citenamefont {Sierou}\ and\ \citenamefont
  {Brady}(2004)}]{sierou2004shear}%
  \BibitemOpen
  \bibfield  {author} {\bibinfo {author} {\bibfnamefont {A.}~\bibnamefont
  {Sierou}}\ and\ \bibinfo {author} {\bibfnamefont {J.~F.}\ \bibnamefont
  {Brady}},\ }\href@noop {} {\bibfield  {journal} {\bibinfo  {journal} {Journal
  of fluid mechanics}\ }\textbf {\bibinfo {volume} {506}},\ \bibinfo {pages}
  {285} (\bibinfo {year} {2004})}\BibitemShut {NoStop}%
\bibitem [{\citenamefont {Pine}\ \emph {et~al.}(2005)\citenamefont {Pine},
  \citenamefont {Gollub}, \citenamefont {Brady},\ and\ \citenamefont
  {Leshansky}}]{pine2005chaos}%
  \BibitemOpen
  \bibfield  {author} {\bibinfo {author} {\bibfnamefont {D.~J.}\ \bibnamefont
  {Pine}}, \bibinfo {author} {\bibfnamefont {J.~P.}\ \bibnamefont {Gollub}},
  \bibinfo {author} {\bibfnamefont {J.~F.}\ \bibnamefont {Brady}}, \ and\
  \bibinfo {author} {\bibfnamefont {A.~M.}\ \bibnamefont {Leshansky}},\
  }\href@noop {} {\bibfield  {journal} {\bibinfo  {journal} {Nature}\ }\textbf
  {\bibinfo {volume} {438}},\ \bibinfo {pages} {997} (\bibinfo {year}
  {2005})}\BibitemShut {NoStop}%
\bibitem [{\citenamefont {Popova}\ \emph {et~al.}(2007)\citenamefont {Popova},
  \citenamefont {Vorobieff}, \citenamefont {Ingber},\ and\ \citenamefont
  {Graham}}]{popova2007interaction}%
  \BibitemOpen
  \bibfield  {author} {\bibinfo {author} {\bibfnamefont {M.}~\bibnamefont
  {Popova}}, \bibinfo {author} {\bibfnamefont {P.}~\bibnamefont {Vorobieff}},
  \bibinfo {author} {\bibfnamefont {M.~S.}\ \bibnamefont {Ingber}}, \ and\
  \bibinfo {author} {\bibfnamefont {A.~L.}\ \bibnamefont {Graham}},\
  }\href@noop {} {\bibfield  {journal} {\bibinfo  {journal} {Physical Review
  E}\ }\textbf {\bibinfo {volume} {75}},\ \bibinfo {pages} {066309} (\bibinfo
  {year} {2007})}\BibitemShut {NoStop}%
\bibitem [{\citenamefont {Blanc}\ \emph {et~al.}(2011)\citenamefont {Blanc},
  \citenamefont {Peters},\ and\ \citenamefont
  {Lemaire}}]{blanc2011experimental}%
  \BibitemOpen
  \bibfield  {author} {\bibinfo {author} {\bibfnamefont {F.}~\bibnamefont
  {Blanc}}, \bibinfo {author} {\bibfnamefont {F.}~\bibnamefont {Peters}}, \
  and\ \bibinfo {author} {\bibfnamefont {E.}~\bibnamefont {Lemaire}},\
  }\href@noop {} {\bibfield  {journal} {\bibinfo  {journal} {Physical review
  letters}\ }\textbf {\bibinfo {volume} {107}},\ \bibinfo {pages} {208302}
  (\bibinfo {year} {2011})}\BibitemShut {NoStop}%
\bibitem [{\citenamefont {Pham}\ \emph {et~al.}(2015)\citenamefont {Pham},
  \citenamefont {Metzger},\ and\ \citenamefont {Butler}}]{pham2015particle}%
  \BibitemOpen
  \bibfield  {author} {\bibinfo {author} {\bibfnamefont {P.}~\bibnamefont
  {Pham}}, \bibinfo {author} {\bibfnamefont {B.}~\bibnamefont {Metzger}}, \
  and\ \bibinfo {author} {\bibfnamefont {J.~E.}\ \bibnamefont {Butler}},\
  }\href@noop {} {\bibfield  {journal} {\bibinfo  {journal} {Physics of
  Fluids}\ }\textbf {\bibinfo {volume} {27}},\ \bibinfo {pages} {051701}
  (\bibinfo {year} {2015})}\BibitemShut {NoStop}%
\bibitem [{\citenamefont {Gallier}\ \emph {et~al.}(2014)\citenamefont
  {Gallier}, \citenamefont {Lemaire}, \citenamefont {Peters},\ and\
  \citenamefont {Lobry}}]{gallier2014rheology}%
  \BibitemOpen
  \bibfield  {author} {\bibinfo {author} {\bibfnamefont {S.}~\bibnamefont
  {Gallier}}, \bibinfo {author} {\bibfnamefont {E.}~\bibnamefont {Lemaire}},
  \bibinfo {author} {\bibfnamefont {F.}~\bibnamefont {Peters}}, \ and\ \bibinfo
  {author} {\bibfnamefont {L.}~\bibnamefont {Lobry}},\ }\href@noop {}
  {\bibfield  {journal} {\bibinfo  {journal} {Journal of Fluid Mechanics}\
  }\textbf {\bibinfo {volume} {757}},\ \bibinfo {pages} {514} (\bibinfo {year}
  {2014})}\BibitemShut {NoStop}%
\bibitem [{\citenamefont {Lyon}\ and\ \citenamefont
  {Leal}(1998)}]{lyon1998experimental}%
  \BibitemOpen
  \bibfield  {author} {\bibinfo {author} {\bibfnamefont {M.}~\bibnamefont
  {Lyon}}\ and\ \bibinfo {author} {\bibfnamefont {L.}~\bibnamefont {Leal}},\
  }\href@noop {} {\bibfield  {journal} {\bibinfo  {journal} {Journal of Fluid
  Mechanics}\ }\textbf {\bibinfo {volume} {363}},\ \bibinfo {pages} {25}
  (\bibinfo {year} {1998})}\BibitemShut {NoStop}%
\bibitem [{\citenamefont {Boyer}\ \emph
  {et~al.}(2011{\natexlab{a}})\citenamefont {Boyer}, \citenamefont
  {Pouliquen},\ and\ \citenamefont {Guazzelli}}]{boyer2011dense}%
  \BibitemOpen
  \bibfield  {author} {\bibinfo {author} {\bibfnamefont {F.}~\bibnamefont
  {Boyer}}, \bibinfo {author} {\bibfnamefont {O.}~\bibnamefont {Pouliquen}}, \
  and\ \bibinfo {author} {\bibfnamefont {{\'E}.}~\bibnamefont {Guazzelli}},\
  }\href@noop {} {\bibfield  {journal} {\bibinfo  {journal} {Journal of Fluid
  Mechanics}\ }\textbf {\bibinfo {volume} {686}},\ \bibinfo {pages} {5}
  (\bibinfo {year} {2011}{\natexlab{a}})}\BibitemShut {NoStop}%
\bibitem [{\citenamefont {Saint-Michel}\ \emph {et~al.}(2019)\citenamefont
  {Saint-Michel}, \citenamefont {Manneville}, \citenamefont {Meeker},
  \citenamefont {Ovarlez},\ and\ \citenamefont {Bodiguel}}]{saint2019x}%
  \BibitemOpen
  \bibfield  {author} {\bibinfo {author} {\bibfnamefont {B.}~\bibnamefont
  {Saint-Michel}}, \bibinfo {author} {\bibfnamefont {S.}~\bibnamefont
  {Manneville}}, \bibinfo {author} {\bibfnamefont {S.}~\bibnamefont {Meeker}},
  \bibinfo {author} {\bibfnamefont {G.}~\bibnamefont {Ovarlez}}, \ and\
  \bibinfo {author} {\bibfnamefont {H.}~\bibnamefont {Bodiguel}},\ }\href@noop
  {} {\bibfield  {journal} {\bibinfo  {journal} {Physics of Fluids}\ }\textbf
  {\bibinfo {volume} {31}},\ \bibinfo {pages} {103301} (\bibinfo {year}
  {2019})}\BibitemShut {NoStop}%
\bibitem [{\citenamefont {d'Ambrosio}\ \emph {et~al.}(2021)\citenamefont
  {d'Ambrosio}, \citenamefont {Blanc},\ and\ \citenamefont
  {Lemaire}}]{dambrosio2021}%
  \BibitemOpen
  \bibfield  {author} {\bibinfo {author} {\bibfnamefont {E.}~\bibnamefont
  {d'Ambrosio}}, \bibinfo {author} {\bibfnamefont {F.}~\bibnamefont {Blanc}}, \
  and\ \bibinfo {author} {\bibfnamefont {E.}~\bibnamefont {Lemaire}},\ }\href
  {\doibase 10.1017/jfm.2020.1074} {\bibfield  {journal} {\bibinfo  {journal}
  {Journal of Fluid Mechanics}\ }\textbf {\bibinfo {volume} {911}},\ \bibinfo
  {pages} {A22} (\bibinfo {year} {2021})}\BibitemShut {NoStop}%
\bibitem [{\citenamefont {Mari}\ \emph {et~al.}(2014)\citenamefont {Mari},
  \citenamefont {Seto}, \citenamefont {Morris},\ and\ \citenamefont
  {Denn}}]{mari2014shear}%
  \BibitemOpen
  \bibfield  {author} {\bibinfo {author} {\bibfnamefont {R.}~\bibnamefont
  {Mari}}, \bibinfo {author} {\bibfnamefont {R.}~\bibnamefont {Seto}}, \bibinfo
  {author} {\bibfnamefont {J.~F.}\ \bibnamefont {Morris}}, \ and\ \bibinfo
  {author} {\bibfnamefont {M.~M.}\ \bibnamefont {Denn}},\ }\href@noop {}
  {\bibfield  {journal} {\bibinfo  {journal} {Journal of Rheology}\ }\textbf
  {\bibinfo {volume} {58}},\ \bibinfo {pages} {1693} (\bibinfo {year}
  {2014})}\BibitemShut {NoStop}%
\bibitem [{\citenamefont {Ch{\`e}vremont}\ \emph {et~al.}(2019)\citenamefont
  {Ch{\`e}vremont}, \citenamefont {Chareyre},\ and\ \citenamefont
  {Bodiguel}}]{chevremont2019quantitative}%
  \BibitemOpen
  \bibfield  {author} {\bibinfo {author} {\bibfnamefont {W.}~\bibnamefont
  {Ch{\`e}vremont}}, \bibinfo {author} {\bibfnamefont {B.}~\bibnamefont
  {Chareyre}}, \ and\ \bibinfo {author} {\bibfnamefont {H.}~\bibnamefont
  {Bodiguel}},\ }\href@noop {} {\bibfield  {journal} {\bibinfo  {journal}
  {Physical Review Fluids}\ }\textbf {\bibinfo {volume} {4}},\ \bibinfo {pages}
  {064302} (\bibinfo {year} {2019})}\BibitemShut {NoStop}%
\bibitem [{\citenamefont {Lhuillier}(2009)}]{lhuillier2009migration}%
  \BibitemOpen
  \bibfield  {author} {\bibinfo {author} {\bibfnamefont {D.}~\bibnamefont
  {Lhuillier}},\ }\href@noop {} {\bibfield  {journal} {\bibinfo  {journal}
  {Physics of Fluids}\ }\textbf {\bibinfo {volume} {21}},\ \bibinfo {pages}
  {023302} (\bibinfo {year} {2009})}\BibitemShut {NoStop}%
\bibitem [{\citenamefont {Nott}\ \emph {et~al.}(2011)\citenamefont {Nott},
  \citenamefont {Guazzelli},\ and\ \citenamefont
  {Pouliquen}}]{nott2011suspension}%
  \BibitemOpen
  \bibfield  {author} {\bibinfo {author} {\bibfnamefont {P.~R.}\ \bibnamefont
  {Nott}}, \bibinfo {author} {\bibfnamefont {E.}~\bibnamefont {Guazzelli}}, \
  and\ \bibinfo {author} {\bibfnamefont {O.}~\bibnamefont {Pouliquen}},\
  }\href@noop {} {\bibfield  {journal} {\bibinfo  {journal} {Physics of
  Fluids}\ }\textbf {\bibinfo {volume} {23}},\ \bibinfo {pages} {043304}
  (\bibinfo {year} {2011})}\BibitemShut {NoStop}%
\bibitem [{\citenamefont {Nott}\ and\ \citenamefont
  {Brady}(1994)}]{nott1994pressure}%
  \BibitemOpen
  \bibfield  {author} {\bibinfo {author} {\bibfnamefont {P.~R.}\ \bibnamefont
  {Nott}}\ and\ \bibinfo {author} {\bibfnamefont {J.~F.}\ \bibnamefont
  {Brady}},\ }\href@noop {} {\bibfield  {journal} {\bibinfo  {journal} {Journal
  of Fluid Mechanics}\ }\textbf {\bibinfo {volume} {275}},\ \bibinfo {pages}
  {157} (\bibinfo {year} {1994})}\BibitemShut {NoStop}%
\bibitem [{\citenamefont {Morris}\ and\ \citenamefont
  {Boulay}(1999)}]{morris1999curvilinear}%
  \BibitemOpen
  \bibfield  {author} {\bibinfo {author} {\bibfnamefont {J.~F.}\ \bibnamefont
  {Morris}}\ and\ \bibinfo {author} {\bibfnamefont {F.}~\bibnamefont
  {Boulay}},\ }\href@noop {} {\bibfield  {journal} {\bibinfo  {journal}
  {Journal of rheology}\ }\textbf {\bibinfo {volume} {43}},\ \bibinfo {pages}
  {1213} (\bibinfo {year} {1999})}\BibitemShut {NoStop}%
\bibitem [{\citenamefont {Leighton}\ and\ \citenamefont
  {Acrivos}(1986)}]{leighton1986viscous}%
  \BibitemOpen
  \bibfield  {author} {\bibinfo {author} {\bibfnamefont {D.}~\bibnamefont
  {Leighton}}\ and\ \bibinfo {author} {\bibfnamefont {A.}~\bibnamefont
  {Acrivos}},\ }\href@noop {} {\bibfield  {journal} {\bibinfo  {journal}
  {Chemical engineering science}\ }\textbf {\bibinfo {volume} {41}},\ \bibinfo
  {pages} {1377} (\bibinfo {year} {1986})}\BibitemShut {NoStop}%
\bibitem [{\citenamefont {Phillips}\ \emph {et~al.}(1992)\citenamefont
  {Phillips}, \citenamefont {Armstrong}, \citenamefont {Brown}, \citenamefont
  {Graham},\ and\ \citenamefont {Abbott}}]{phillips1992constitutive}%
  \BibitemOpen
  \bibfield  {author} {\bibinfo {author} {\bibfnamefont {R.~J.}\ \bibnamefont
  {Phillips}}, \bibinfo {author} {\bibfnamefont {R.~C.}\ \bibnamefont
  {Armstrong}}, \bibinfo {author} {\bibfnamefont {R.~A.}\ \bibnamefont
  {Brown}}, \bibinfo {author} {\bibfnamefont {A.~L.}\ \bibnamefont {Graham}}, \
  and\ \bibinfo {author} {\bibfnamefont {J.~R.}\ \bibnamefont {Abbott}},\
  }\href@noop {} {\bibfield  {journal} {\bibinfo  {journal} {Physics of Fluids
  A: Fluid Dynamics}\ }\textbf {\bibinfo {volume} {4}},\ \bibinfo {pages} {30}
  (\bibinfo {year} {1992})}\BibitemShut {NoStop}%
\bibitem [{\citenamefont {Acrivos}\ \emph {et~al.}(1993)\citenamefont
  {Acrivos}, \citenamefont {Mauri},\ and\ \citenamefont
  {Fan}}]{acrivos1993shear}%
  \BibitemOpen
  \bibfield  {author} {\bibinfo {author} {\bibfnamefont {A.}~\bibnamefont
  {Acrivos}}, \bibinfo {author} {\bibfnamefont {R.}~\bibnamefont {Mauri}}, \
  and\ \bibinfo {author} {\bibfnamefont {X.}~\bibnamefont {Fan}},\ }\href@noop
  {} {\bibfield  {journal} {\bibinfo  {journal} {International journal of
  multiphase flow}\ }\textbf {\bibinfo {volume} {19}},\ \bibinfo {pages} {797}
  (\bibinfo {year} {1993})}\BibitemShut {NoStop}%
\bibitem [{\citenamefont {Snook}\ \emph {et~al.}(2016)\citenamefont {Snook},
  \citenamefont {Butler},\ and\ \citenamefont {Guazzelli}}]{snook2016dynamics}%
  \BibitemOpen
  \bibfield  {author} {\bibinfo {author} {\bibfnamefont {B.}~\bibnamefont
  {Snook}}, \bibinfo {author} {\bibfnamefont {J.~E.}\ \bibnamefont {Butler}}, \
  and\ \bibinfo {author} {\bibfnamefont {{\'E}.}~\bibnamefont {Guazzelli}},\
  }\href@noop {} {\bibfield  {journal} {\bibinfo  {journal} {Journal of Fluid
  Mechanics}\ }\textbf {\bibinfo {volume} {786}},\ \bibinfo {pages} {128}
  (\bibinfo {year} {2016})}\BibitemShut {NoStop}%
\bibitem [{\citenamefont {Zarraga}\ \emph {et~al.}(2000)\citenamefont
  {Zarraga}, \citenamefont {Hill},\ and\ \citenamefont
  {Leighton~Jr}}]{zarraga2000characterization}%
  \BibitemOpen
  \bibfield  {author} {\bibinfo {author} {\bibfnamefont {I.~E.}\ \bibnamefont
  {Zarraga}}, \bibinfo {author} {\bibfnamefont {D.~A.}\ \bibnamefont {Hill}}, \
  and\ \bibinfo {author} {\bibfnamefont {D.~T.}\ \bibnamefont {Leighton~Jr}},\
  }\href@noop {} {\bibfield  {journal} {\bibinfo  {journal} {Journal of
  Rheology}\ }\textbf {\bibinfo {volume} {44}},\ \bibinfo {pages} {185}
  (\bibinfo {year} {2000})}\BibitemShut {NoStop}%
\bibitem [{\citenamefont {Boyer}\ \emph
  {et~al.}(2011{\natexlab{b}})\citenamefont {Boyer}, \citenamefont
  {Guazzelli},\ and\ \citenamefont {Pouliquen}}]{boyer2011unifying}%
  \BibitemOpen
  \bibfield  {author} {\bibinfo {author} {\bibfnamefont {F.}~\bibnamefont
  {Boyer}}, \bibinfo {author} {\bibfnamefont {{\'E}.}~\bibnamefont
  {Guazzelli}}, \ and\ \bibinfo {author} {\bibfnamefont {O.}~\bibnamefont
  {Pouliquen}},\ }\href@noop {} {\bibfield  {journal} {\bibinfo  {journal}
  {Physical Review Letters}\ }\textbf {\bibinfo {volume} {107}},\ \bibinfo
  {pages} {188301} (\bibinfo {year} {2011}{\natexlab{b}})}\BibitemShut
  {NoStop}%
\bibitem [{\citenamefont {Schowalter}\ \emph {et~al.}(1968)\citenamefont
  {Schowalter}, \citenamefont {Chaffey},\ and\ \citenamefont
  {Brenner}}]{Brenner}%
  \BibitemOpen
  \bibfield  {author} {\bibinfo {author} {\bibfnamefont {W.}~\bibnamefont
  {Schowalter}}, \bibinfo {author} {\bibfnamefont {C.}~\bibnamefont {Chaffey}},
  \ and\ \bibinfo {author} {\bibfnamefont {H.}~\bibnamefont {Brenner}},\
  }\href@noop {} {\bibfield  {journal} {\bibinfo  {journal} {Journal of colloid
  and interface science}\ }\textbf {\bibinfo {volume} {26}},\ \bibinfo {pages}
  {152} (\bibinfo {year} {1968})}\BibitemShut {NoStop}%
\bibitem [{\citenamefont {Ramachandran}\ \emph {et~al.}(2010)\citenamefont
  {Ramachandran}, \citenamefont {Loewenberg},\ and\ \citenamefont
  {Leighton~Jr}}]{ramachandran2010constitutive}%
  \BibitemOpen
  \bibfield  {author} {\bibinfo {author} {\bibfnamefont {A.}~\bibnamefont
  {Ramachandran}}, \bibinfo {author} {\bibfnamefont {M.}~\bibnamefont
  {Loewenberg}}, \ and\ \bibinfo {author} {\bibfnamefont {D.~T.}\ \bibnamefont
  {Leighton~Jr}},\ }\href@noop {} {\bibfield  {journal} {\bibinfo  {journal}
  {Physics of Fluids}\ }\textbf {\bibinfo {volume} {22}},\ \bibinfo {pages}
  {083301} (\bibinfo {year} {2010})}\BibitemShut {NoStop}%
\bibitem [{\citenamefont {King}\ and\ \citenamefont
  {Leighton~Jr}(2001)}]{king2001measurement}%
  \BibitemOpen
  \bibfield  {author} {\bibinfo {author} {\bibfnamefont {M.~R.}\ \bibnamefont
  {King}}\ and\ \bibinfo {author} {\bibfnamefont {D.~T.}\ \bibnamefont
  {Leighton~Jr}},\ }\href@noop {} {\bibfield  {journal} {\bibinfo  {journal}
  {Physics of Fluids}\ }\textbf {\bibinfo {volume} {13}},\ \bibinfo {pages}
  {397} (\bibinfo {year} {2001})}\BibitemShut {NoStop}%
\bibitem [{\citenamefont {Hollingsworth}\ and\ \citenamefont
  {Johns}(2006)}]{hollingsworth2006droplet}%
  \BibitemOpen
  \bibfield  {author} {\bibinfo {author} {\bibfnamefont {K.}~\bibnamefont
  {Hollingsworth}}\ and\ \bibinfo {author} {\bibfnamefont {M.}~\bibnamefont
  {Johns}},\ }\href@noop {} {\bibfield  {journal} {\bibinfo  {journal} {Journal
  of colloid and interface science}\ }\textbf {\bibinfo {volume} {296}},\
  \bibinfo {pages} {700} (\bibinfo {year} {2006})}\BibitemShut {NoStop}%
\bibitem [{\citenamefont {Abbas}\ \emph {et~al.}(2017)\citenamefont {Abbas},
  \citenamefont {Pouplin}, \citenamefont {Masbernat}, \citenamefont
  {Lin{\'e}},\ and\ \citenamefont {D{\'e}carre}}]{abbas2017pipe}%
  \BibitemOpen
  \bibfield  {author} {\bibinfo {author} {\bibfnamefont {M.}~\bibnamefont
  {Abbas}}, \bibinfo {author} {\bibfnamefont {A.}~\bibnamefont {Pouplin}},
  \bibinfo {author} {\bibfnamefont {O.}~\bibnamefont {Masbernat}}, \bibinfo
  {author} {\bibfnamefont {A.}~\bibnamefont {Lin{\'e}}}, \ and\ \bibinfo
  {author} {\bibfnamefont {S.}~\bibnamefont {D{\'e}carre}},\ }\href@noop {}
  {\bibfield  {journal} {\bibinfo  {journal} {AIChE Journal}\ }\textbf
  {\bibinfo {volume} {63}},\ \bibinfo {pages} {5182} (\bibinfo {year}
  {2017})}\BibitemShut {NoStop}%
\bibitem [{\citenamefont {Kosvintsev}\ \emph {et~al.}(2005)\citenamefont
  {Kosvintsev}, \citenamefont {Gasparini}, \citenamefont {Holdich},
  \citenamefont {Cumming},\ and\ \citenamefont
  {Stillwell}}]{kosvintsev2005liquid}%
  \BibitemOpen
  \bibfield  {author} {\bibinfo {author} {\bibfnamefont {S.~R.}\ \bibnamefont
  {Kosvintsev}}, \bibinfo {author} {\bibfnamefont {G.}~\bibnamefont
  {Gasparini}}, \bibinfo {author} {\bibfnamefont {R.~G.}\ \bibnamefont
  {Holdich}}, \bibinfo {author} {\bibfnamefont {I.~W.}\ \bibnamefont
  {Cumming}}, \ and\ \bibinfo {author} {\bibfnamefont {M.~T.}\ \bibnamefont
  {Stillwell}},\ }\href@noop {} {\bibfield  {journal} {\bibinfo  {journal}
  {Industrial \& engineering chemistry research}\ }\textbf {\bibinfo {volume}
  {44}},\ \bibinfo {pages} {9323} (\bibinfo {year} {2005})}\BibitemShut
  {NoStop}%
\bibitem [{\citenamefont {Maleki}\ \emph {et~al.}(2021)\citenamefont {Maleki},
  \citenamefont {de~Loubens}, \citenamefont {Xie}, \citenamefont {Talansier},
  \citenamefont {Bodiguel},\ and\ \citenamefont
  {Leonetti}}]{maleki2021membrane}%
  \BibitemOpen
  \bibfield  {author} {\bibinfo {author} {\bibfnamefont {M.}~\bibnamefont
  {Maleki}}, \bibinfo {author} {\bibfnamefont {C.}~\bibnamefont {de~Loubens}},
  \bibinfo {author} {\bibfnamefont {K.}~\bibnamefont {Xie}}, \bibinfo {author}
  {\bibfnamefont {E.}~\bibnamefont {Talansier}}, \bibinfo {author}
  {\bibfnamefont {H.}~\bibnamefont {Bodiguel}}, \ and\ \bibinfo {author}
  {\bibfnamefont {M.}~\bibnamefont {Leonetti}},\ }\href@noop {} {\bibfield
  {journal} {\bibinfo  {journal} {Chemical Engineering Science}\ }\textbf
  {\bibinfo {volume} {237}},\ \bibinfo {pages} {116567} (\bibinfo {year}
  {2021})}\BibitemShut {NoStop}%
\bibitem [{\citenamefont {Bentley}\ and\ \citenamefont
  {Leal}(1986)}]{bentley1986experimental}%
  \BibitemOpen
  \bibfield  {author} {\bibinfo {author} {\bibfnamefont {B.}~\bibnamefont
  {Bentley}}\ and\ \bibinfo {author} {\bibfnamefont {L.~G.}\ \bibnamefont
  {Leal}},\ }\href@noop {} {\bibfield  {journal} {\bibinfo  {journal} {Journal
  of Fluid Mechanics}\ }\textbf {\bibinfo {volume} {167}},\ \bibinfo {pages}
  {241} (\bibinfo {year} {1986})}\BibitemShut {NoStop}%
\bibitem [{\citenamefont {Henschke}\ \emph {et~al.}(2002)\citenamefont
  {Henschke}, \citenamefont {Schlieper},\ and\ \citenamefont
  {Pfennig}}]{henschke2002determination}%
  \BibitemOpen
  \bibfield  {author} {\bibinfo {author} {\bibfnamefont {M.}~\bibnamefont
  {Henschke}}, \bibinfo {author} {\bibfnamefont {L.~H.}\ \bibnamefont
  {Schlieper}}, \ and\ \bibinfo {author} {\bibfnamefont {A.}~\bibnamefont
  {Pfennig}},\ }\href@noop {} {\bibfield  {journal} {\bibinfo  {journal}
  {Chemical Engineering Journal}\ }\textbf {\bibinfo {volume} {85}},\ \bibinfo
  {pages} {369} (\bibinfo {year} {2002})}\BibitemShut {NoStop}%
\bibitem [{\citenamefont {Aouane}\ \emph {et~al.}(2021)\citenamefont {Aouane},
  \citenamefont {Scagliarini},\ and\ \citenamefont {Harting}}]{Aouane}%
  \BibitemOpen
  \bibfield  {author} {\bibinfo {author} {\bibfnamefont {O.}~\bibnamefont
  {Aouane}}, \bibinfo {author} {\bibfnamefont {A.}~\bibnamefont {Scagliarini}},
  \ and\ \bibinfo {author} {\bibfnamefont {J.}~\bibnamefont {Harting}},\
  }\href@noop {} {\bibfield  {journal} {\bibinfo  {journal} {Journal of Fluid
  Mechanics}\ }\textbf {\bibinfo {volume} {911}} (\bibinfo {year}
  {2021})}\BibitemShut {NoStop}%
\bibitem [{\citenamefont {Richardson}\ and\ \citenamefont
  {Zaki}(1997)}]{richardson1997sedimentation}%
  \BibitemOpen
  \bibfield  {author} {\bibinfo {author} {\bibfnamefont {J.}~\bibnamefont
  {Richardson}}\ and\ \bibinfo {author} {\bibfnamefont {W.}~\bibnamefont
  {Zaki}},\ }\href@noop {} {\bibfield  {journal} {\bibinfo  {journal} {Chemical
  Engineering Research and Design}\ }\textbf {\bibinfo {volume} {75}},\
  \bibinfo {pages} {S82} (\bibinfo {year} {1997})}\BibitemShut {NoStop}%
\bibitem [{\citenamefont {Zinchenko}(1984)}]{zinchenko1984effect}%
  \BibitemOpen
  \bibfield  {author} {\bibinfo {author} {\bibfnamefont {A.}~\bibnamefont
  {Zinchenko}},\ }\href@noop {} {\bibfield  {journal} {\bibinfo  {journal}
  {Journal of Applied Mathematics and Mechanics}\ }\textbf {\bibinfo {volume}
  {48}},\ \bibinfo {pages} {198} (\bibinfo {year} {1984})}\BibitemShut
  {NoStop}%
\end{thebibliography}%


\begin{thebibliography}{3}%
\makeatletter
\providecommand \@ifxundefined [1]{%
 \@ifx{#1\undefined}
}%
\providecommand \@ifnum [1]{%
 \ifnum #1\expandafter \@firstoftwo
 \else \expandafter \@secondoftwo
 \fi
}%
\providecommand \@ifx [1]{%
 \ifx #1\expandafter \@firstoftwo
 \else \expandafter \@secondoftwo
 \fi
}%
\providecommand \natexlab [1]{#1}%
\providecommand \enquote  [1]{``#1''}%
\providecommand \bibnamefont  [1]{#1}%
\providecommand \bibfnamefont [1]{#1}%
\providecommand \citenamefont [1]{#1}%
\providecommand \href@noop [0]{\@secondoftwo}%
\providecommand \href [0]{\begingroup \@sanitize@url \@href}%
\providecommand \@href[1]{\@@startlink{#1}\@@href}%
\providecommand \@@href[1]{\endgroup#1\@@endlink}%
\providecommand \@sanitize@url [0]{\catcode `\\12\catcode `\$12\catcode
  `\&12\catcode `\#12\catcode `\^12\catcode `\_12\catcode `\%12\relax}%
\providecommand \@@startlink[1]{}%
\providecommand \@@endlink[0]{}%
\providecommand \url  [0]{\begingroup\@sanitize@url \@url }%
\providecommand \@url [1]{\endgroup\@href {#1}{\urlprefix }}%
\providecommand \urlprefix  [0]{URL }%
\providecommand \Eprint [0]{\href }%
\providecommand \doibase [0]{https://doi.org/}%
\providecommand \selectlanguage [0]{\@gobble}%
\providecommand \bibinfo  [0]{\@secondoftwo}%
\providecommand \bibfield  [0]{\@secondoftwo}%
\providecommand \translation [1]{[#1]}%
\providecommand \BibitemOpen [0]{}%
\providecommand \bibitemStop [0]{}%
\providecommand \bibitemNoStop [0]{.\EOS\space}%
\providecommand \EOS [0]{\spacefactor3000\relax}%
\providecommand \BibitemShut  [1]{\csname bibitem#1\endcsname}%
\let\auto@bib@innerbib\@empty
\bibitem [{\citenamefont {Kosvintsev}\ \emph {et~al.}(2005)\citenamefont
  {Kosvintsev}, \citenamefont {Gasparini}, \citenamefont {Holdich},
  \citenamefont {Cumming},\ and\ \citenamefont
  {Stillwell}}]{kosvintsev2005liquid}%
  \BibitemOpen
  \bibfield  {author} {\bibinfo {author} {\bibfnamefont {S.~R.}\ \bibnamefont
  {Kosvintsev}}, \bibinfo {author} {\bibfnamefont {G.}~\bibnamefont
  {Gasparini}}, \bibinfo {author} {\bibfnamefont {R.~G.}\ \bibnamefont
  {Holdich}}, \bibinfo {author} {\bibfnamefont {I.~W.}\ \bibnamefont
  {Cumming}},\ and\ \bibinfo {author} {\bibfnamefont {M.~T.}\ \bibnamefont
  {Stillwell}},\ }\href@noop {} {\bibfield  {journal} {\bibinfo  {journal}
  {Industrial \& engineering chemistry research}\ }\textbf {\bibinfo {volume}
  {44}},\ \bibinfo {pages} {9323} (\bibinfo {year} {2005})}\BibitemShut
  {NoStop}%
\bibitem [{\citenamefont {Henschke}\ \emph {et~al.}(2002)\citenamefont
  {Henschke}, \citenamefont {Schlieper},\ and\ \citenamefont
  {Pfennig}}]{henschke2002determination}%
  \BibitemOpen
  \bibfield  {author} {\bibinfo {author} {\bibfnamefont {M.}~\bibnamefont
  {Henschke}}, \bibinfo {author} {\bibfnamefont {L.~H.}\ \bibnamefont
  {Schlieper}},\ and\ \bibinfo {author} {\bibfnamefont {A.}~\bibnamefont
  {Pfennig}},\ }\href@noop {} {\bibfield  {journal} {\bibinfo  {journal}
  {Chemical Engineering Journal}\ }\textbf {\bibinfo {volume} {85}},\ \bibinfo
  {pages} {369} (\bibinfo {year} {2002})}\BibitemShut {NoStop}%
\bibitem [{\citenamefont {Saint-Michel}\ \emph {et~al.}(2019)\citenamefont
  {Saint-Michel}, \citenamefont {Manneville}, \citenamefont {Meeker},
  \citenamefont {Ovarlez},\ and\ \citenamefont {Bodiguel}}]{saint2019x}%
  \BibitemOpen
  \bibfield  {author} {\bibinfo {author} {\bibfnamefont {B.}~\bibnamefont
  {Saint-Michel}}, \bibinfo {author} {\bibfnamefont {S.}~\bibnamefont
  {Manneville}}, \bibinfo {author} {\bibfnamefont {S.}~\bibnamefont {Meeker}},
  \bibinfo {author} {\bibfnamefont {G.}~\bibnamefont {Ovarlez}},\ and\ \bibinfo
  {author} {\bibfnamefont {H.}~\bibnamefont {Bodiguel}},\ }\href@noop {}
  {\bibfield  {journal} {\bibinfo  {journal} {Physics of Fluids}\ }\textbf
  {\bibinfo {volume} {31}},\ \bibinfo {pages} {103301} (\bibinfo {year}
  {2019})}\BibitemShut {NoStop}%
\end{thebibliography}%

\end{document}


\title{Supplementary material for:\\
Resuspension of droplets}

\author{Mehdi Maleki}
 \author{Clément de Loubens}
  \email{clement.de-loubens@univ-grenoble-alpes.fr}
\author{Hugues Bodiguel}

 \affiliation{Univ. Grenoble Alpes, CNRS, Grenoble INP, LRP, 38000 Grenoble, France}

\date{\today}

\maketitle
\tableofcontents
\section{Droplet production}

The  oil-in-water emulsions were produced with two different methods: microfluidics and membrane emulsification. Monodisperse emulsions were fabricated by microfluidics method while polydisperse emulsions were produced by membrane emulsification. The aqueous phase  consisted of water-glycerol mixture (glycerol 84 \% w/w, CAS number 56-81-5, VWR) and added food grade colorant (0.001\% w/w, E122, Breton). The oil phase was composed of medium chain triglyceride oil (Nestlé, Switzerland) with sorbitan  trioleate  85  (1\% w/w, CAS  number  26266-58-0,  Sigma-Aldrich). Table \ref{table_production} shows the physical properties of oil and water phases of the emulsions.

\begin{table}[h!]
  \begin{center}
    \caption{Physical properties of oil and water phases of the emulsions at 23 $^{\circ}$C.}
    \label{table_production}
    \newcommand\T{\rule{0pt}{2.6ex}} 
    \newcommand\B{\rule[-1.2ex]{0pt}{0pt}} 
    \addtolength{\tabcolsep}{3pt} 
    \begin{tabular}{c c c c}
    \hline
      Phase  & \makecell{Density \T\\ kg.m$^{-3}$ \B} & \makecell{Viscosity\\ mPa.s} & \makecell{Refractive Index\\nD} \\
      \hline
      Oil Phase  \T\B & 943.42 & 50 & 1.44948 \\
      \hline
      Water Phase \T\B & 1218.52 & 85 & 1.44948 \\
      \hline
    \end{tabular}
  \end{center}
\end{table}

\begin{figure*}
    \centering
    \includegraphics[width=0.7\textwidth]{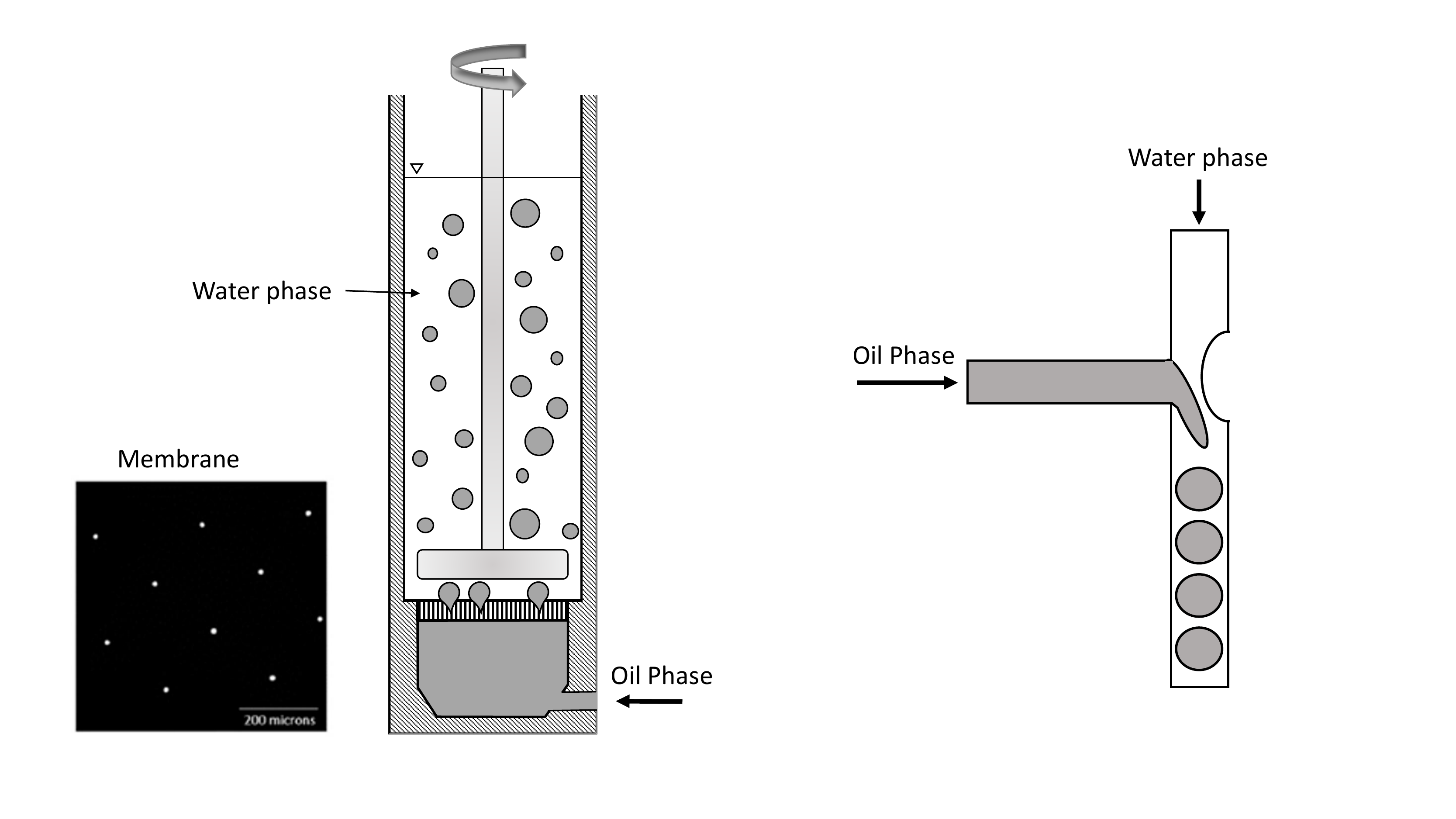}
    \caption{Production of droplet emulsions. \textbf{Left:} Membrane emulsification system to produce poly dispersed droplets, \textbf{Right:} Microfluidic T-junction chip to produce monodispesed droplets}
    \label{production}
\end{figure*}

Monodisperse oil-in-water  emulsions were produced with a custom-made PMMA T-junction microfluidic chip with a cross-section 1$\times$1 mm. The inside of the chip was treated with acetone in order to render the surface hydrophilic. As shown in figure \ref{production}-right, the continuous aqueous phase was injected through the main channel while the dispersed oil phase was introduced through the perpendicular branch via a round glass capillary (CM Scientific Ltd) of 300 $\mu$m inner diameter. The flow rates were controlled by two syringe pumps (neMSYS, CETONI). To improve the production rate, T-junction chip was modified by narrowing the intersection area to strength shear forces. Two monodisperse emulsions with different droplet sizes were generated at volumetric frow rates of oil phase 0.2 and 0.3 mL/min while the flow rate of aqueous phase 2 mL /min in both cases. 

The lab-scale membrane emulsification system was purchased from  Micropore Technologies Ltd (UK) under the commercial name Micropore LDC-1. The system includes a micro-engineered emulsification membrane with 20 $\mu$m  cylindrical laser etched pores  under a  paddle-blade stirrer (Figure \ref{production}-left). The rotational velocity of the stirrer was controlled by a DC motor. The thin flat nickel emulsification membrane  was chemically treated on one side to have a hydrophilic surface. The pore spacing and porosity of the membrane were 200 $\mu$m and 0.91\% respectively. The array of pores was located in an  narrow annular region on the membrane to limit the variation of shear rate and thus homogenize the droplet size distribution \cite{kosvintsev2005liquid}. The aqueous phase was stirred with the paddle at 500 rpm and the oil phase was injected by a syringe pump (neMSYS, CETONI) through the membrane with a flow rate of 2 ml/min. 

\section{Size Distribution}

\begin{figure}
    \centering
    \includegraphics[width= \columnwidth]{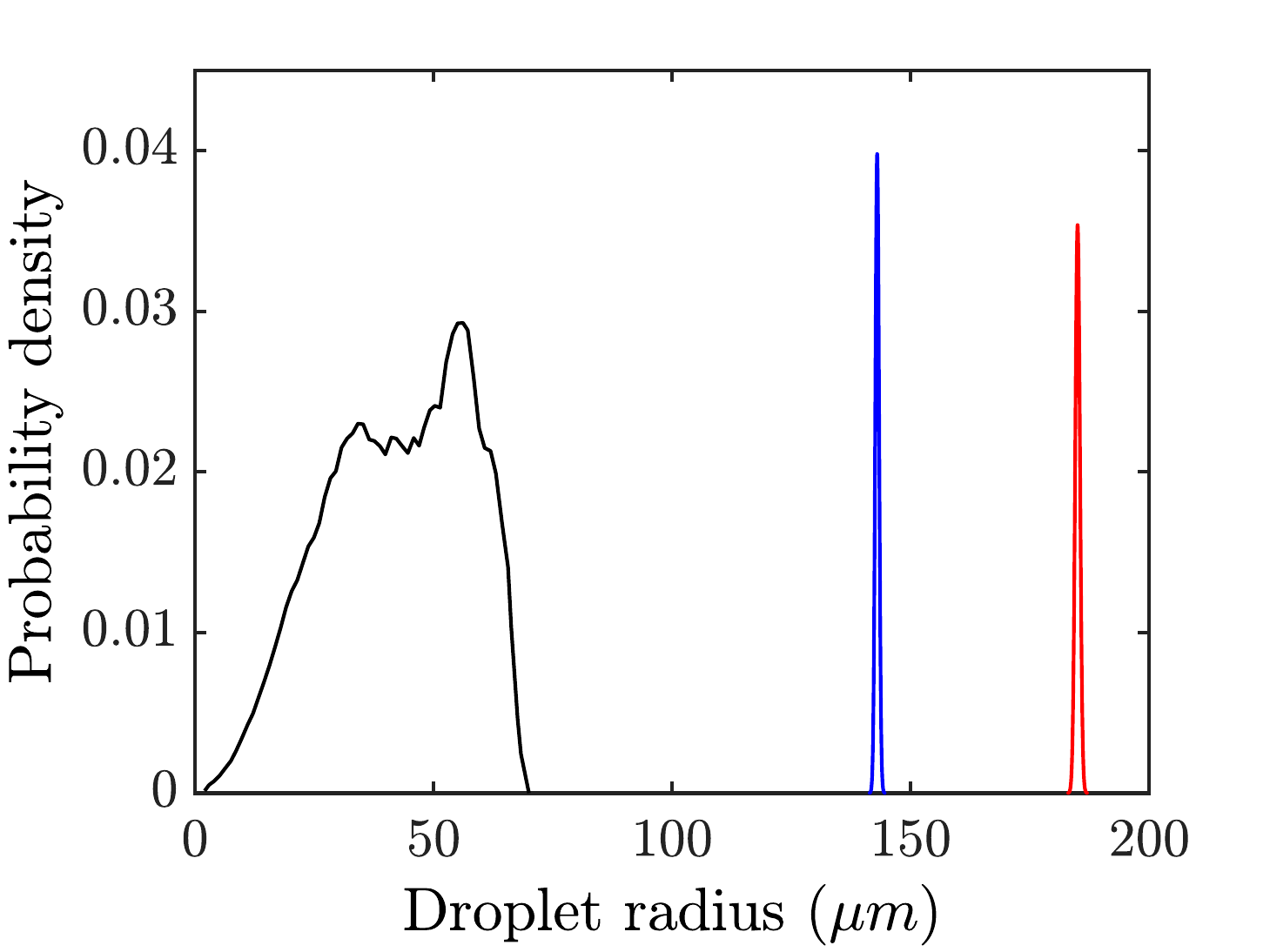}
    \caption{Probability density of volumic size distribution of droplets. Polydisperse emulsion was resulted from membrane emulsification system and monodisperse emulsion was produced by microfluidic system.}
    \label{size}
\end{figure}

The estimation of the droplet size distribution was based on bright field microscopy images. A large amount of droplets was placed on wide microscope slides and put under an inverted microscope (IX-73, Olympus) equipped with 4, 10, and 20 fold objectives. The motorized stage (Marzhauser) was automatized to perform a tile scan on the entire specimen and the pictures were acquired with  a digital camera (ORCA-Flash 4.0, Hamamatsu). The  droplet size was measured with a custom written software based on the Matlab (The MathWorks, Inc.) image processing toolbox.  The average radius of the monodisperse droplets were $a = 143$ and $185\, \mu$m and polydisperse droplets $a = 47 \, \mu$m with a sampling rate over several thousand. The size distribution of the monodisperse and polydisperse emulsions is shown in Figure \ref{size}.  

\section{Parallax issue}

\begin{figure*}
    \centering
    \includegraphics[width= \textwidth]{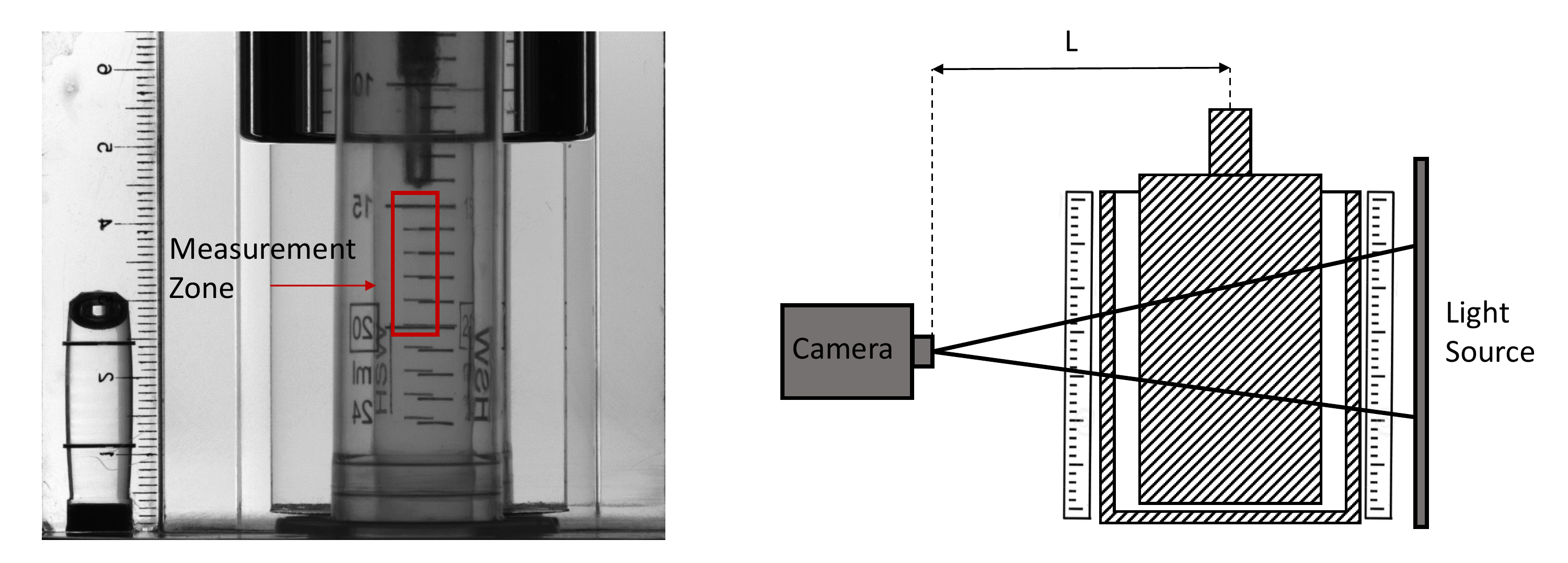}
    \caption{Parallax issue along the vertical direction solved by increasing the distance $L$ between the camera and Taylor-Couette cell up to five meters. To have high resolution pictures of measurement zone a lens with a large focal length has been used. A ruler behind and front of the Taylor-Couette cell indicates a negligible parallax issue.}
    \label{parallax}
\end{figure*}

In this section, we analyze the parallax issue raised by the finite distance between the camera and Taylor-Couette geometry. In the experimental setup, the light source was a 2D LED panel with a homogeneous and stable light intensity. It was placed parallel to the Taylor-Couette geometry with 10 cm of distance. As shown in figure \ref{parallax}-right, light beams traveled through the suspension and geometry, then they were collected by a camera (Basler, acA2500-14gc), located at a distance of L from the geometry. 

By considering a finite value of L, the camera received the incident light beams within an angle of 0 to $\alpha$ relative to the horizontal direction. This lead to a smoothing effect on the concentration profile in the vertical direction $x_3$ through a length of $\textit{l}$ (figure \ref{parallax}-right). This means that the measurement of the droplet concentration was vertically averaged in this length. Consequently, the errors induced by smoothing could be considerable where the vertical concentration sharply shifted to zero at the nose of the suspension as shown in figure \ref{profile_poly}. Note that, measuring of normal viscosity $\eta_{n,3}$ and determining exponent $n$ in SBM model depends strongly on the  precision to detect the curvature of the concentration transition to zero. 

By increasing the distance L between the camera and the geometry, the angle $\alpha$ and the vertical averaging length $\textit{l}$ reduced. Consequently, the spatial smoothing of  the concentration profile in the vertical direction  became less effective. The experiments were performed with a distance of L up to 5 m, whereas the height of the suspension in the measurement zone was 2 cm. To have high resolution images (5 pixels for a droplet diameter) we used a lens with a large focal length. Figure \ref{parallax}-left represents an image of the geometry with two similar rulers placed at two sides of the geometry. We can observe that the parallax in the measurement zone was negligible. However it became more decisive near the bottom of the geometry. In such configuration of the experimental setup, we estimated $\alpha = 0.11$ degree and vertical averaging length $\textit{l}=96 \, \mu$m. The ratio of $\textit{l}$ to the diameter of monodisperse and polydisperse droplets is $\textit{l}/a = 1/3$ and $1$ respectively. These small ratios indicated a negligible effect of the parallax on the concentration profile measurement where it experienced a sharp transition to zero. Consequently, the parallax did not impact the comparison of our measurements with SBM model.
\section{Tuning refractive index}

\begin{figure}
    \centering
    \includegraphics[width= \columnwidth]{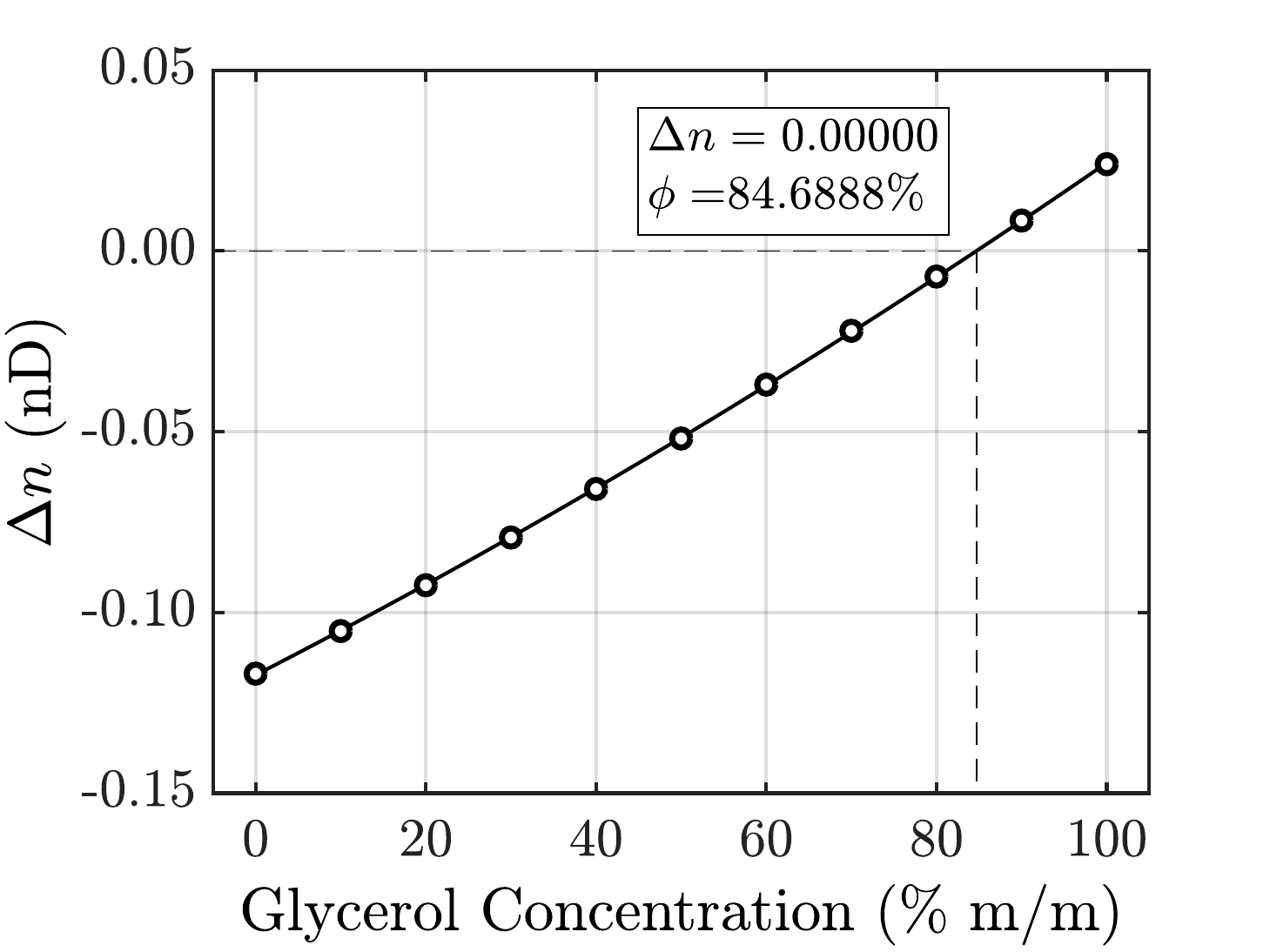}
    \caption{The difference of refractive index between the droplets and suspending fluid as a function of glycerol volume fraction at 23 $^{\circ}$C and wavelength $\lambda = 589$ nm. A zero contrast of refractive index was estimated with glycerol concentration $\phi = 84.6888 \%$. The final contrast was refined to a precision of $\Delta n = 0.00000$ nD by adding  some drops of glycerol or water.}
    \label{index}
\end{figure}

Balancing the refractive index between the droplets and suspending fluid was  crucial for our experimental method. Light refraction between the two phases in the emulsions could give rise to considerable errors in the measurement of the concentration profile $\phi(x_3)$. The refractive index of the oil phase was measured with  precise refractometer (Abbemat  350,  Anton  Paar)  $n = 1.44948$ nD at  $T = 23 \ ^{\circ}$C and wavelength $\lambda = 589$ nm. Note that the resuspension experiments were conducted at the same temperature and a wavelength band of $\lambda = 525 \pm 15$ nm. Figure \ref{index} illustrates the difference  of refractive index between the droplets and suspending fluid for various glycerol volume fractions in the suspending fluid. The zero contrast was estimated with a glycerol volume fraction around $\phi = 84.68 \%$ w/w. Then after preparing the solution, the contrast was refined to a precision of $\Delta n = 0.00000$ nD with drop-by-drop addition of glycerol or water and mixed.  After that 1 mL of the solution was pipetted  from different cites in the solution volume. The refractive index of these samples was measured to the precision $\Delta n = 0.00000$ nD and in the case of mismatch the process was repeated. 

\section{Characterizing the absorbance}

\begin{figure*}
    \centering
    \includegraphics[width= \textwidth]{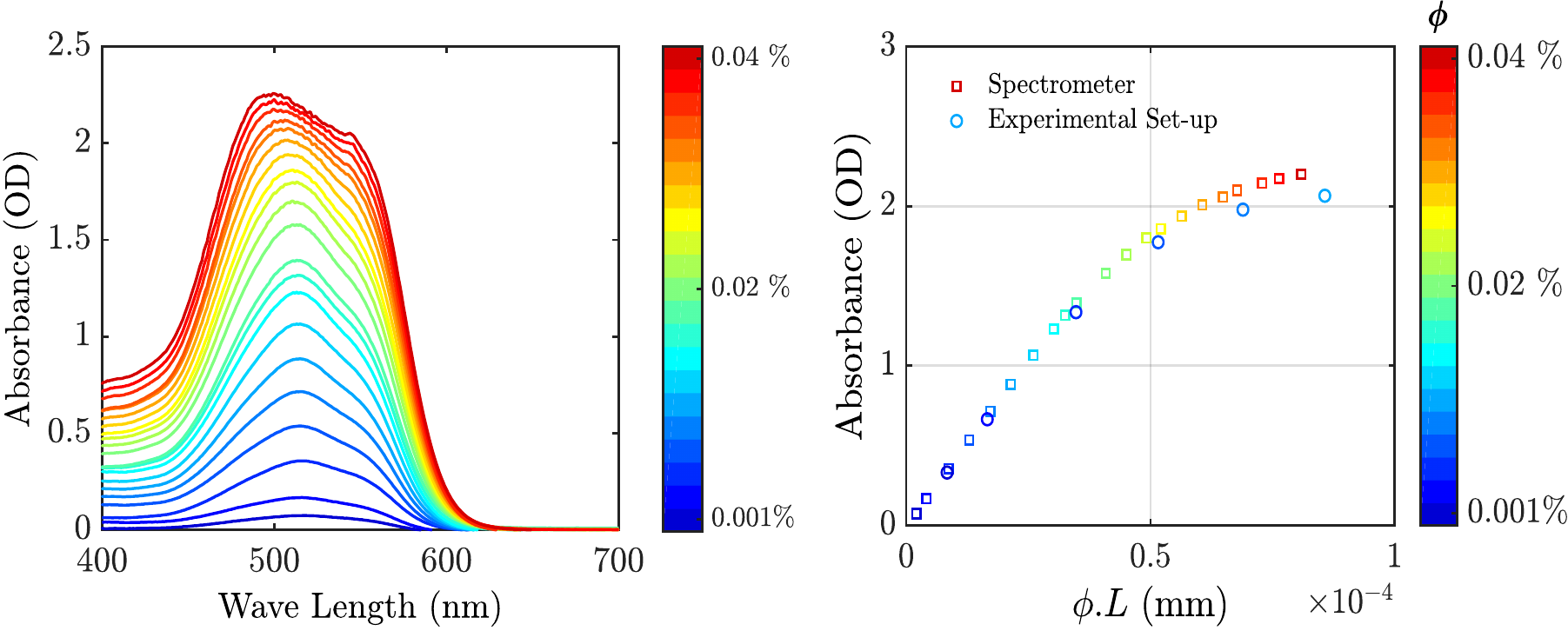}
    \caption{\small Characterization of the light absorbance of the colorant, \textbf{Left:} The light absorbance of the colorant for concentration ranging from $c = 0.001\%$ to $0.04\%$ mg/mL in visible light wave length measured by a spectrometer in a $L=2$ mm cuvette, \textbf{Right:} Absorbance of the colorant at wave length $\lambda = 525$ nm as a function of $\phi L$, where $\phi$ is the colorant concentration and $L$ is the traveled length of light in the sample. The measured absorbance by the spectrometer in a cuvette with $L = 2$ mm has an excellent agreement in the linear part with measured absorbance by the experimental set-up in the Taylor-Couette cell. A bandpass filter with the  center wavelength, $\lambda = 525 \pm 15$nm was used in the experimental set-up.}
    \label{absorbance}
\end{figure*}

The vertical concentration profile of the resuspended emulsions $\phi(x_3)$ was obtained by the light absorption technique. Since the droplets and suspending fluid were both completely transparent, adding a colorant to the suspending fluid provided a contrast in the light absorbance degree between the two phases. A non-fluorescent food-grade additive, E122 (Breton) was used as the colorant. UV-Vis spectrophotometry analysis of the colorant was performed in a quartz cuvette (Hellma Analytics) with a light path of $L=2$ mm and a portable spectrometer (RedLite, Ocean Insight). Figure \ref{absorbance}-left demonstrates the absorption spectra of the colorant dissolved in the suspending fluid with the concentration ranging from  $c = 0.001\%$ to $0.04\%$ mg/ml. The absorption spectrum was a broad band with a maximum absorbance peak at $\lambda = 520$ nm. 

To validate our experimental approach, we measured the absorbance of the colorant with a digital camera (acA2500-14gc, Basler) in  Taylor-Couette Cell and compared the results with a spectrophotometer. As shown in figure \ref{absorbance}-left, the wavelength related to the peak of absorbance was around $\lambda = 520$ nm, thus we used an interference bandpass filter with the  center wavelength  $\lambda = 525 \pm 15$ nm. Beer-Lambert law formulates the exponential decay of the light intensity passing through a solution. Thus we related intensities and a distance L and calculated the absorbance as:
\begin{equation}
A \equiv log(\frac{I_0 - I_d}{I_L-I_d}) = \varepsilon c L
\end{equation}
where $I_0$ is the light intensity after traveling the suspending fluid without colorant as well as the transparent container cell of sample, $I_L$ is the traveled light intensity with contribution of the colorant, $I_d$ is the measured light intensity by the insulated camera which represents the noise and $\varepsilon$ is the attenuation rate of light for the colorant. Contrary to the cuvette, in Taylor-Couette cell, the traveled length of the light beam through the sample was not constant. By determining the corresponding traveled length $L$ for each pixel of the captured image, a 2D light absorbance map was obtained. As expected, for the homogeneous colorant solution in Taylor-Couette cell, we obtained an uniform 2D absorbance map. Figure \ref{absorbance}-right demonstrates the absorbance of the colorant at wavelength $\lambda = 520$ nm, measured by the spectrophotometer and the experimental setup as a function of $\phi .L$ where $\phi$ is the colorant concentration and $L$ is the traveled length of light in the sample. The measurements show an excellent consistency in the linear regime. The experiments were conducted in such fashion to keep the corresponding absorbance values within the linear part.

\section{Coalescence issue}
\label{section_coalescence}

\begin{figure*}
    \centering
    \includegraphics[width= \textwidth]{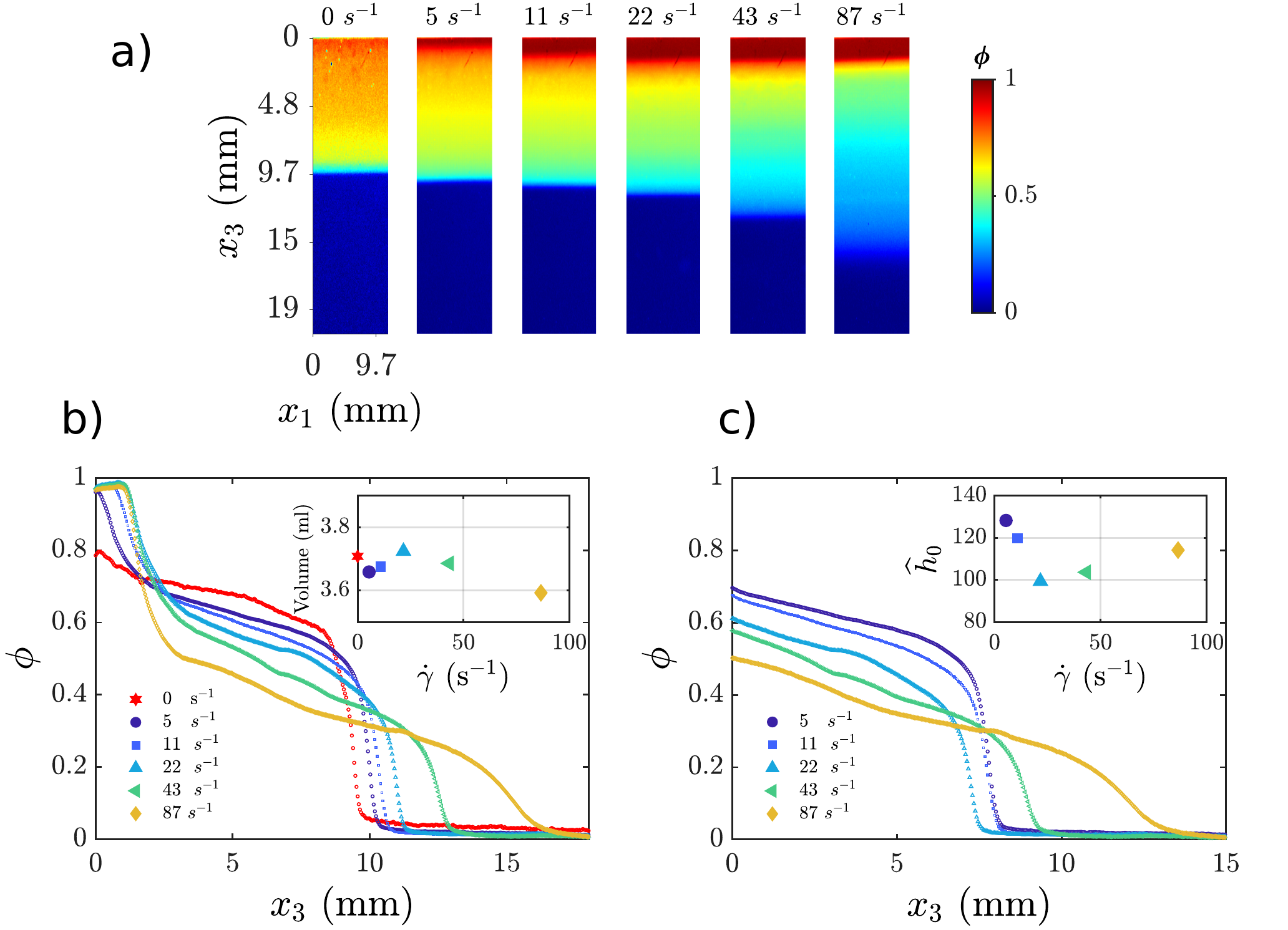}
    \caption{Coalescence issue for polydisperse suspensions despite surface treatment, \textbf{a:} 2D concentration profile of droplets in $x_1-x_3$ plane and presence of an oil layer at the top of the Taylor-Couette cell resulted from coalescence of droplets, \textbf{b:} The concentration profile of droplets in steady state along the resuspension direction $x_3$ for shear rates ranging from $\dot{\gamma} = 0$ to $87\ s^{-1}$. The coalesced part was detected by sharp increase of the concentration. Insert figure is the total volume of oil phase calculated by equation \ref{eq_volume}. \textbf{c:} Corrected concentration profile of droplets as a function of $x_3$. Inset figure is normalized height of the creamed suspension at rest $\widehat{h}_0 = h/a$ calculated by the equation \ref{eq_h0_coalescence} as a function of shear rate. $h$ is the height of resuspension in the steady state and $a$ is the average size of the droplets.}
    \label{coalescence}
\end{figure*}

As the droplets were less dense than the suspending fluid, they creamed on  top of the Taylor-Couette cell. The droplets could be deformed under the buoyant force, which causes them to coalesce \citep{henschke2002determination}.  The droplet coalescence was inhibited with  a surface treatment of the geometry for monodisperse suspensions. However, this surface treatment proved to be inefficient for polydisperse suspensions. Figure \ref{coalescence}-a shows the  concentration map of the polydisperse droplets in $x_1-x_3$ plane for shear rates ranging from $\dot{\gamma} = 0$ to $87\ s^{-1}$ in steady state. It illustrates the emergence and then stabilization of the oil layer caused by coalescence at the top of the geometry.  Additionally, we observed that the creamed suspension underwent a similar transformation over a short period of time even at rest. However, under a shear stress, this process slowed down. We had to make a  correction in the experimental data in order to discard the coalesced part and took into account only the true quantity of droplets in the suspension. This was necessary since the migration phenomenon in the resuspension experiments depended on the droplet quantity. Moreover, comparing the migration rate in the same suspension as a function of shear rate was possible only if the non-coalesced droplet quantity was taken into account. 

\begin{figure*}
    \centering
    \includegraphics[width= 0.95\textwidth]{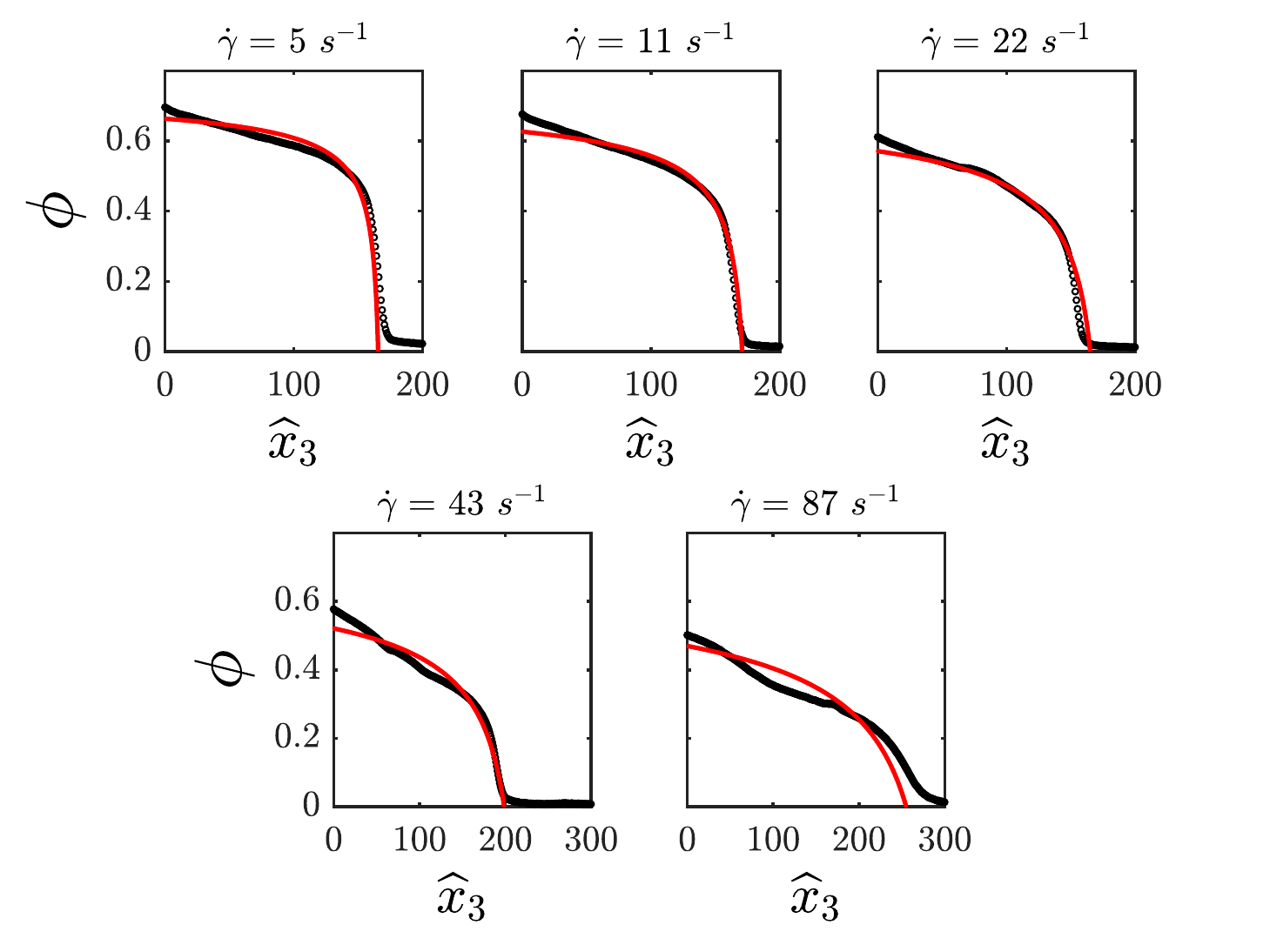}
    \caption{Corrected concentration profile of polydisperse suspension of droplets for shear rates in the range of $\dot{\gamma} = 5$ to $87 \ s^{-1}$. The circles are the experimental data from section \ref{section_coalescence} and solid lines are the analytical profiles based on the Suspension Balance Model with $n=2$ and calculated from equation \ref{Eq_phix3} with $\phi_m = 0.76$ and $\lambda_3 = 0.69$.}
    \label{profile_poly}
\end{figure*}

Figure \ref{coalescence}-b shows the vertical steady state concentration profile of droplets $\phi(x_3)$ for shear rates ranging from $\dot{\gamma} = 0$ to $87\ s^{-1}$ . The total volume of the oil phase (droplets and coalesced part) was calculated by integrating $\phi(x_3)$ over $x_3$:
\begin{equation}
    V = \pi (R_2^2 - R_1^2) \int_{0}^{H} \phi(x_3) \,dx_3
    \label{eq_volume}
\end{equation}
where $R_1 = 20$, $R_2 = 24$ and $H = 50$ mm are the inner, outer radii and height of the Taylor-Couette cell respectively. The total volume of oil phase (inset figure \ref{coalescence}-b) was V$=3.691 \pm 0.033$ ml for shear rates up to $\dot{\gamma} = 43 \ s^{-1}$ which corresponds to a relative standard deviation from  2.7 \%  up to = 5.2 \% at $\dot{\gamma} = 87 \ s^{-1}$. This variation increment originated from the optical issues at the surface of oil layer with the air. The coalesced part was detected by a sharp increase in the concentration profile near $x_3 = 0$. Figure \ref{coalescence}-c shows the same concentration profiles after eliminating the part related to the coalescence. Consequently, each concentration profile was a representation of the droplets migration as a response to the shear rate but with different quantities of particles in the suspension. Based on the mass conservation, by integrating the concentration profile over the normalized length $\widehat{x}_3 = x_3/a$, we estimated the normalized height of the creamed suspension at rest $\widehat{h}_0$ for each concentration profile:
\begin{equation}
    \widehat{h}_0 \phi_m = \int_{0}^{\widehat{h}} \phi(\widehat{x}_3) \,d\widehat{x}_3
    \label{eq_h0_coalescence}
\end{equation}
where $\widehat{h}$ is the normalized height of the suspension in steady state, and took $\phi_m = 0.76$. The inset of figure \ref{coalescence}-c shows the calculated $\widehat{h}_0$ for each concentration profile which is in the range of  100 to 128 . The increase of $\widehat{h}_0$ for shear rates $\dot{\gamma} = 43$ and $87 \ s^{-1}$ can be attributed to the break-up of the large oil pockets which did not coalesce with the oil layer located at the top of the geometry. To compare our results, such as the normal viscosity $\eta_{n,3}$, the evolution of suspension height $\widehat{h}$ and concentration profile $\phi(\widehat{x}_3)$, to Suspension Balance Model (SBM) we used the corrected data.

\section{Concentration profiles at rest}

At rest and after creaming during about 12h, the volume fraction was not uniform but exhibited a gradient from about 0.6 at the bottom of the emulsion layer up to 0.8 at the top. This effect was the signature of droplet deformability, since - at least for monodisperse systems -  a volume fraction of 0.8 cannot be obtained with spherical droplets. When comparing the particle stress, given by $\Sigma_{p,33} = \int \Delta \rho g \phi dx_3$ with the Laplace pressure, we indeed found that the particle stress was a non negligible fraction of Laplace pressure, which allowed significant deformation of the droplet. 

\begin{figure}
	\centering
	\includegraphics[width=1\columnwidth]{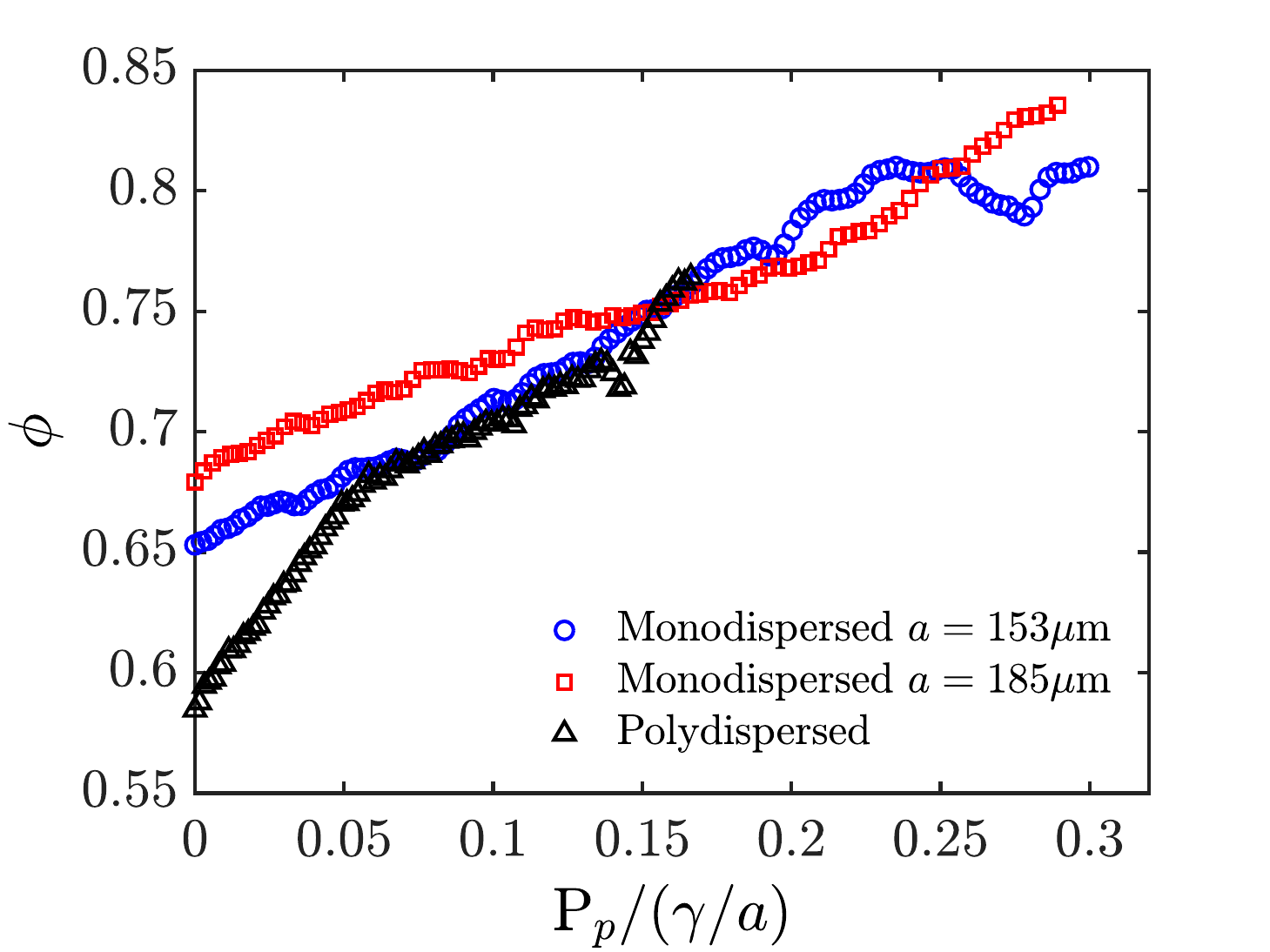}
	\caption{Concentration profiles at rest ($\dot{\gamma}=0$) as a function of the Laplace number $L_A$ for monodispersed and polydispersed suspenions of droplets.}
	\label{Fig_Laplace}
\end{figure}

In Fig. \ref{Fig_Laplace}, the volume fraction at rest is plotted for the three systems studied as a function of the Laplace number, $La= \Sigma_{p,33}/(\gamma/a)$. The two monodisperse systems nicely collapsed on a master curve, which was roughly an affine function of $La$. This validated the above hypothesis that the droplet deform due to the buoyant mass of the layer underneath. The polydisperse system slighlty deviated from this behaviour in the bottom of the layer, which might indicate that a size segregation existed in this case.    

\section{Concentration profile of polydisperse suspension}

Based on the theoretical framework of the Suspension Balance Model (SBM), in the case of $n=2$, the concentration profile of the droplets was calculated analytically as \citep{saint2019x}:

\begin{equation}
    \frac{\phi(\widehat{x}_3)}{\phi_m} = 1 - \left[ 1 + \frac{\phi_m}{\lambda_3Sh}(\widehat{h}-\widehat{x}_3)\right]^{-1/2} 
    \label{Eq_phix3}
\end{equation}

where the normalized steady state height of the suspension   $\widehat{h}$ is
\begin{equation}
  \widehat{h} = \widehat{h}_0 + 2 \sqrt{\frac{\lambda_3 \widehat{h}_0}{\phi_m}Sh} 
  \label{eq_height}
\end{equation}
Figure \ref{profile_poly} represents the corrected concentration profiles of the polydisperse suspension and the analytical profile calculated by equation \ref{Eq_phix3} for shear rates ranging from $\dot{\gamma} = 5$ to $87 \ s^{-1}$. Jamming concentration was $\phi_m = 0.76$ and the best fit was found with free parameter $\lambda_3 = 0.69$. The experimental results for shear rates $\dot{\gamma} = 5$ to $43 \ s^{-1}$ were in good agreement with the theoretical prediction in equation \ref{Eq_phix3}. In the case of $\dot{\gamma} = 87\ s^{-1}$ the inertia effects came into play. Furthermore, the discrepancies between the experimental and theoretical results could  originated from the structuration of the droplets caused by size polydispersity.

\begin{figure}[tb!]
    \centering
    \includegraphics[width= \columnwidth]{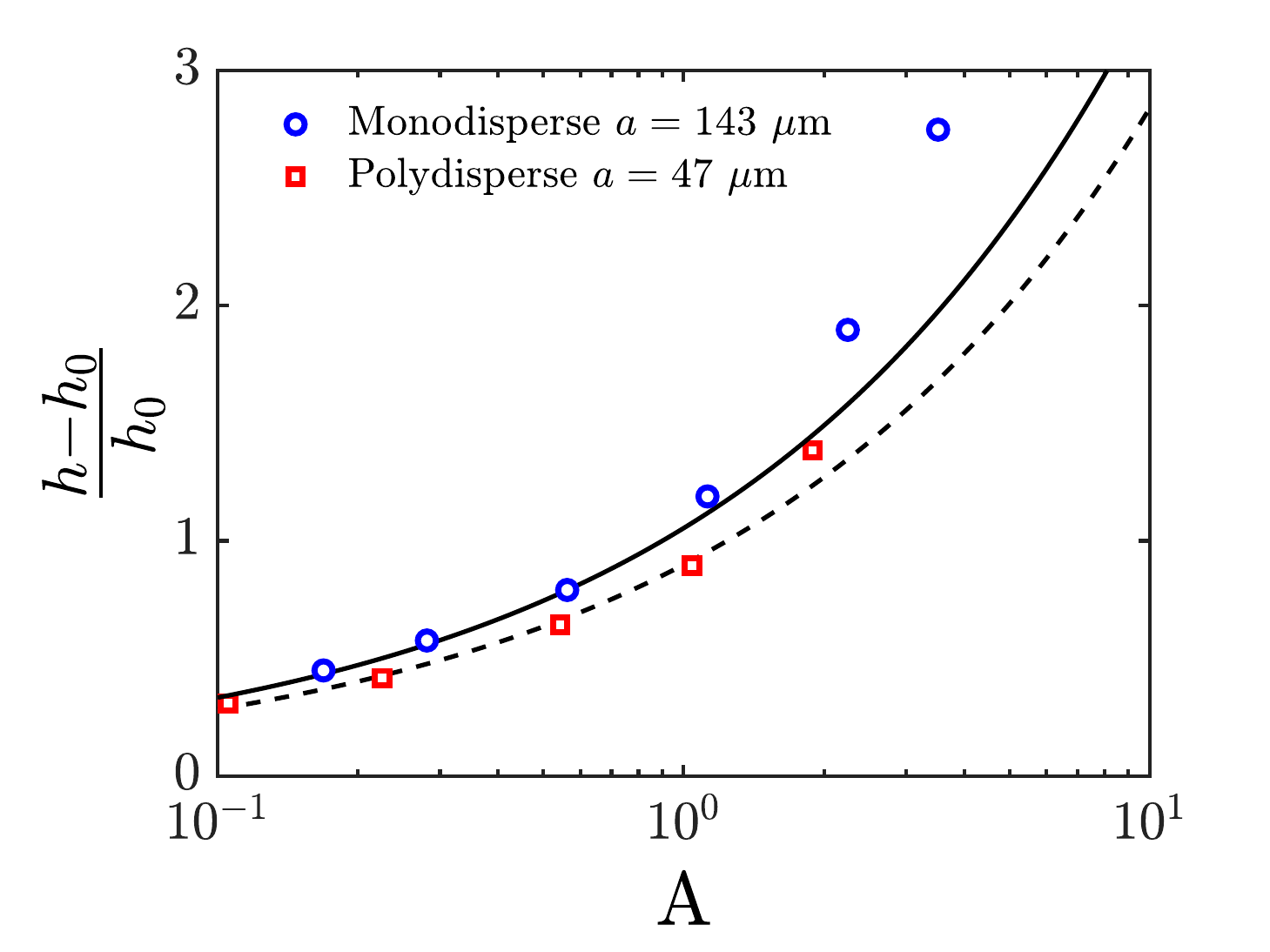}
    \caption{Variation of the normalize height of resuspended monodisperse and polydisperse suspensions as a function of Acrivos number. Solid line is the correlation with $n=2$, $\phi_m = 0.8$ and $\lambda_3 = 1$ for monodisperse droplets and dashed line with $n=2$, $\phi_m = 0.76$ and $\lambda_3 = 0.69$ for monodisperse and polydisperse droplets respectively.}
    \label{fig_height}
\end{figure}

\section{Suspension height evolution}

Normalized height of the suspensions for shear rates in the range of $\dot{\gamma} = 5$ to $135 s^{-1}$  is presented in figure \ref{fig_height} as a function of Acrivos number. Acrivos number is defined: 
\begin{equation}
    A=\frac{\eta_0\dot{\gamma}}{\Delta \rho g h_0}
\end{equation}
which is the ratio of viscous force to buoyancy force as Shields number $Sh$ but with $h_0$ for the characteristic length. The analytical correlation presented in equation \ref{eq_height} is traced for monodisperse droplets (solid line) with $n=2$, $\phi_m = 0.8$ and $\lambda_3 = 1$ and for polydisperse droplets (dashed line) with $n=2$, $\phi_m = 0.76$ and $\lambda_3 = 0.69$. We observed that the experimental results for both monodisperse and polydisperse suspensions were in excellent agreement with the theoretical correlation in the range of $0.1<A<1$. In contrast, a divergence was found for higher values of $A$ which can be due to the inertial effects.

\bibliography{refs}